\documentclass[iop,apj,twocolappendix,numberedappendix,revtex4]{emulateapj}	

\usepackage[breaklinks,colorlinks,citecolor=blue]{hyperref}		
\usepackage{enumerate}
\usepackage{color}      
\usepackage{amsmath}
\usepackage[thickspace,squaren]{SIunits} 

\usepackage[nolist]{acronym}

\usepackage{epsfig}

\usepackage[normalem]{ulem}	
\def\mathbi#1{\textbf{\em #1}}
\newcommand{\erf}{\ensuremath{\text{erf}}} 

\renewcommand{\vec}[1]{\ensuremath{\mathbi{#1}}} 
\newcommand{\gvec}[1]{\ensuremath{\mbox{\boldmath$ #1 $}}} 
\newcommand{\uvec}[1]{\ensuremath{\hat{\mathbi{#1}}}} 
\newcommand{\dprod}{\ensuremath{\pmb{\cdot}}} 
\newcommand{\cprod}{\ensuremath{\pmb{\times}}} 
\newcommand{\abs}[1]{\left| #1 \right|} 
\newcommand{\pd}[2]{\frac{\partial #1}{\partial #2}} 
\newcommand{\pdd}[2]{\frac{\partial^2 #1}{\partial #2^2}} 
\newcommand{\grad}[1]{\gvec{\nabla} #1} 
\renewcommand{\div}[1]{\gvec{\nabla} \dprod #1} 
\newcommand{\curl}[1]{\gvec{\nabla} \cprod #1} 
\providecommand{\abs}[1]{\lvert#1\rvert}

\newcommand{\ps}{\reciprocal\second}
\newcommand{\mps}{\metre\:\ps}      
\newcommand{\mmps}{\metre\squared\:\ps}      
\newcommand{\cmmps}{\centi\mmps}      
\newcommand{\kmmps}{\kilo\mmps}      
\newcommand{\radps}{\radian\:\ps}      
\newcommand{\gauss}{\text{G}}
\newcommand{\maxwell}{\text{Mx}}
\newcommand{\years}{\text{years}}
\renewcommand{\phm}{\phantom}
\newcommand{\tblmk}{\tablenotemark}

\newcommand{\8}{\infty}

\shorttitle{$2\times2$D Babcock--Leighton dynamo. II. Dynamo solutions}
\shortauthors{A. Lemerle \& P. Charbonneau}

\begin{document}


\submitted{Draft version 2016 October 30; published in ApJ 2017 January 9: \url{http://doi.org/10.3847/1538-4357/834/2/133}}

\title{A coupled $2\times2$D Babcock--Leighton solar dynamo model. II. Reference dynamo solutions}


\author{Alexandre Lemerle\altaffilmark{1,2}}
\author{Paul Charbonneau\altaffilmark{1}}

\altaffiltext{1}{D\'epartement de physique, Universit\'e de Montr\'eal, 
2900 boul. \'Edouard-Montpetit, Montr\'eal, QC, H3T 1J4, Canada;
lemerle@astro.umontreal.ca, paulchar@astro.umontreal.ca}
\altaffiltext{2}{Coll\`ege de Bois-de-Boulogne, 10555 av. 
Bois-de-Boulogne, Montr\'eal, QC, H4N 1L4, Canada.}


\begin{abstract}
In this paper we complete the presentation of a new hybrid $2\times2$D 
flux transport dynamo (FTD) model of the solar cycle based on the 
Babcock--Leighton mechanism of poloidal magnetic field regeneration 
via the surface decay of bipolar magnetic regions (BMRs). 
This hybrid model is constructed by allowing the surface flux transport 
(SFT) simulation described in \citet{Lemerle2015} to provide the 
poloidal source term to an axisymmetric FTD simulation defined in a 
meridional plane, which in turn generates the BMRs required by the SFT. 
A key aspect of this coupling is the definition of an emergence function 
describing the probability of BMR emergence as a function of 
the spatial distribution of the internal axisymmetric magnetic field. 
We use a genetic algorithm to calibrate this function, together 
with other model parameters, against observed cycle 21 emergence data. 
We present a reference dynamo solution reproducing many solar cycle 
characteristics, including good hemispheric coupling, phase relationship 
between the surface dipole and the BMR-generating internal field, 
and correlation between dipole strength at cycle maximum and peak 
amplitude of the next cycle. 
The saturation of the cycle amplitude takes place through the quenching 
of the BMR tilt as a function of the internal field. 
The observed statistical scatter about the mean BMR tilt, built into 
the model, acts as a source of stochasticity which dominates amplitude 
fluctuations. 
The model thus can produce Dalton-like epochs of strongly suppressed 
cycle amplitude lasting a few cycles and can even shut off entirely 
following an unfavorable sequence of emergence events. 
\end{abstract}


\section{Introduction}

Close to a century has now gone by since the discovery of the underlying
magnetic nature of the eleven-year sunspot cycle \citep{Hale1919}. 
The magnetic polarity reversals of the leading and following (with 
respect to rotation) components of large \acp{BMR} is now 
thought to reflect the presence, somewhere in the solar interior, of a 
large-scale, dominantly axisymmetric zonally-oriented (toroidal) 
magnetic field, antisymmetric about the sun's equator and itself 
undergoing polarity reversals approximately every eleven years, for a 
full magnetic cycle period of $\simeq22$~years. 
The rotational shear of a pre-existing dipole, later detected on the 
solar surface \citep{Babcock1955}, can act as an inductive source for 
such an internal toroidal magnetic flux system. 
However, closing the dynamo loop requires an inductive mechanism capable 
of regenerating the dipole from this internal toroidal component, in a 
manner such as leading the cyclic polarity reversals of both of these 
large-scale components of the solar magnetic field.

Many candidates for this toroidal-to-poloidal hydromagnetic inductive 
mechanisms have been identified, starting with cyclonic convection 
\citep{Parker1955} and its associated mean electromotive force, and the 
surface decay of bipolar magnetic regions \citep{Babcock1961}, now 
referred to as the \ac{BL} mechanism. 
These were joined more recently by helical waves along thin magnetic 
flux tubes (\citealt{Schmitt1987}; \citealt{Ossendrijver2000}, and 
references therein), and shear instabilities in the tachocline 
\citep{Dikpati2001}, the stably stratified rotational shear layer 
located beneath the base of the solar convection zone, as revealed by 
helioseismology.
In all cases, the rotational influence mediated by the Coriolis force
is the key agent that breaks the mirror symmetry of the inductive 
flows, thus allowing to circumvent Cowling's theorem.

Of these various candidates for poloidal field regeneration, the 
\ac{BL} mechanism stands out as the only one that can be 
directly observed operating at the solar surface, and as such is far 
better constrained than any other. 
In particular, the distribution of tilt angles of \acp{BMR}, namely the 
angle defined by a line segment joining each pole of the \ac{BMR} 
measured with respect to the east--west direction, is now well 
characterized from white light \citep{Howard1991,DasiEspuig2010} and 
magnetographic observations \citep{Wang1989-0}. 
This tilt arises through the action of the Coriolis force,
and associated with it is a net dipole moment so that, effectively, a
poloidal magnetic component is being produced from the pre-existing 
deep-seated toroidal component ultimately giving rise to emerging BMRs
(see \citealt{Fan2009} for a review).
The magnitude of this tilt, and its pattern of variations with 
latitude, \ac{BMR} flux and separation, and statistical fluctuations 
about the mean, all play a key role in setting the magnitude of the 
surface dipole moment produced in the course of a sunspot cycle.

Because the \ac{BL} mechanism operates at the solar surface, 
a transport mechanism is also needed to carry the surface poloidal 
magnetic field down into the interior, where rotational shearing is 
taking place. 
Here again a number of appropriate candidate mechanisms are available, 
including advection by large-scale meridional flows pervading the solar 
convection zone, as well as turbulent transport effects, namely 
isotropic diffusive transport and directional turbulent pumping. 
Viewed globally, the \ac{BL} mechanism is a non-local 
inductive effect: the surface source of poloidal field is driven by the 
deep-seated toroidal component, on timescales much shorter than the 
magnetic cycle period.

Dynamo models of the solar cycle relying on the \ac{BL} 
mechanism of poloidal field regeneration have undergone a vigorous 
revival in the last 25 years or so, spurred by \citet{Wang1989}, 
\citet{Wang1991}, \citet{Choudhuri1995}, and \citet{Durney1995}. 
Many such models are now dispersed in the literature
(for recent reviews see \citealt{Charbonneau2010}; \citealt{Karak2014}).
The vast majority rely on a two-dimensional axisymmetric formulation of 
the problem, whereby the large-scale flows and magnetic field components 
are both axisymmetric, and the dynamo equations solved in a meridional 
$(r,\theta)$ plane. 
Typically, helioseismology-compatible parameterizations for solar-like 
internal differential rotation and meridional circulation are 
introduced, and these flows are assumed steady (the so-called kinematic 
approximation).

Many such models do differ in how they incorporate the \ac{BL}
mechanism, a fundamentally non-axisymmetric effect, into the 
axisymmetric dynamo equations (compare, e.g.,
\citealt{Durney1995,Dikpati1999,Nandy2001,Munoz2010}). 
They also differ in assumptions made regarding the primary magnetic 
field transport mechanism. 
As a consequence, models based on rather different input physics can do 
roughly as well as one another in reproducing the primary 
characteristics of the observed solar cycle. 
However, the differences can matter a lot in practice. 
Perhaps no better illustration of this point can be found than the 
widely differing dynamo model-based predictions of sunspot cycle 24 
made by \cite{Dikpati2006a} and \cite{Choudhuri2007}, each using a 
distinct \ac{BL} model ``calibrated'' to earlier sunspot cycles.

This problem is compounded when introducing data assimilation into the 
model-based prediction, as the datasets must then also be preprocessed
in some way to accommodate the axisymmetric formulation of the dynamo 
model used for forecasting. 
Both aforecited model-based prediction schemes used distinct 
geometrically simplified implementations of different datasets being 
assimilated, and in all likelihood these differences also contributed 
to the widely varying predictions for the amplitude of cycle 24.
Ideally, data assimilation should be carried out using full-disk
magnetograms and/or detailed observations of active region emergences, 
including complete positional and timing information. 
Either way, this requires a dynamo model with a geometrically complete 
representation of the solar surface, and thus demands abandoning 
axisymmetry.

One extreme possibility consists in turning to global 
magnetohydrodynamical simulations of solar convection. 
Despite remarkable progress in the past decade (for a review see, e.g., 
\S~3 of \citealt{Charbonneau2014}), such simulations still cannot 
accommodate sufficient spatial resolution to resolve convection and 
magnetic field evolution in the surface layers, or even capture the 
interior process of magnetic flux rope formation and buoyant rise 
(but on the latter do see \citealt{FanFang2014}; 
\citealt{Nelson2013}; \citeyear{Nelson2014}). 
Typically, such simulations also fail to drive regular, solar-like 
polarity cyclic reversals in the large-scale magnetic field they 
generate (see \citealt{Passos2014b} for the closest thing yet).

Intermediate approaches are also possible: finding a way to include the 
full non-axisymmetric representation of, at-least, the surface 
processes, while retaining the kinematic approach for the transport of 
magnetic flux. 
To our knowledge, only two such models exist in the literature 
\citep[hereafter MD2014]{YeatesMunoz2013,Miesch2014}, as they include a full 
three-dimensional kinematic representation of the solar convection zone 
up to the surface.
Here again, they mostly differ in how they incorporate the localized 
emergence of new magnetic flux: 
\citet{YeatesMunoz2013} impose localized flow perturbations at the base 
of the convection zone to trigger the eruption of active regions out of 
the toroidal flux, while MD2014 and \citet{Miesch2016} 
apply a surface flux deposition technique, through an empirical masking 
of the deep-seated toroidal field. 

In this series of paper we present a \ac{BL} dynamo model 
that belongs to this same category. 
We retain a fairly conventional two-dimensional axisymmetric kinematic 
\ac{FTD} model, specifically the model described in 
\citet{Charbonneau2005}, without its non-local poloidal source term, 
and couple it to a two-dimensional \ac{SFT} simulation. 
The latter provides the source term for the former through the upper 
boundary condition, and in turn the \ac{FTD} provides the emergences 
required as input to the \ac{SFT} simulation. 
We opted to call this a ``$2\times2$D'' dynamo model. 
This is still a kinematic model, in that it uses steady parametrized 
large-scale flow fields compatible with helioseismology and surface 
measurements. 
Specifying the form of these flows requires the adjustment of many model 
parameters, in order to generate the most ``solar-like'' dynamo 
solutions possible.

In \citet[hereafter Paper~I]{Lemerle2015} we introduced a 
genetic algorithm-based 
method for formally carrying out this optimization 
problem, in the context of the surface flux transport simulation. 
The optimization process is set to minimize deviations with 
respect to synoptic magnetograms (and derived global quantities). 
Not only does this approach finds an optimal solution, but it 
also allows to map a range of acceptable solutions, thus providing 
robust Monte Carlo-like confidence intervals on best-fit model 
parameters and allowing the identification of parameter degeneracies.
A key result is that the range of acceptable surface meridional flow 
profiles fits nicely surface Doppler measurements \citep{Ulrich2010}, 
even though these data are not used to constrain the optimization 
process.

In the present paper we extend the procedure to the coupled model 
described above, and thus produce an ``optimal'' $2\times2$D \ac{BL} 
dynamo model of the solar cycle. 
The use of quotes is motivated by the fact that even this basic optimal 
model involves unavoidable stochastic components, associated with the 
flux emergence process, so that it can only fit the Sun (meaning, e.g., 
the sunspot number time series) in a statistical sense.
Indeed, the \ac{SFT} solutions presented in Paper~I already show how 
the uncertainties in global cycle characteristics are dominated by the 
inherent stochasticity of the flux emergence process.

In \S~2 we discuss the formulation of the coupled model and its 
components. 
In \S~3 we turn to its calibration against observed solar 
features. 
In \S~4 we present self-consistent reference dynamo solutions and 
examine their patterns of long term variability.
In \S~5 we discuss the limitations of the calibration technique and 
compare some of the results with direct solar observations.
We conclude by summarizing our most salient results as well as possible 
paths of improvement and ongoing work.

\section{Model} \label{s_model}

The contemporary version of the original scenario proposed by 
\citet{Babcock1959} runs as follows: 
\begin{enumerate}[(i)]
\item[(0)] at solar maximum, strong toroidal magnetic fields are present 
deep in the solar interior, antisymmetric with respect to the equator; 
\item during the ascending and descending phases of the solar cycle, 
toroidal flux loops rise and emerge at the solar surface in the form 
of \acp{BMR}, twisted due to the Coriolis effect, such that the western 
spots tend to be closer to the equator (tilt following on average 
Joy's law); 
\item surface diffusion/transport near the equator allows for more 
cancellation of the western polarities, when merging with their 
counterparts from the other hemisphere, leaving the remaining 
``eastern'' flux to be transported toward the poles and trigger the 
polarity reversal of magnetic polar caps; 
\item the new surface dipole is subducted and sheared by differential 
rotation, building up a new internal toroidal magnetic structure, 
opposite to the preceding one and ready for...
\item ...the generation of a new population of \acp{BMR} during the next 
half-cycle
(from now on, we refer to such half magnetic cycle, or sunspot cycle, 
as simply a ``cycle'').
\end{enumerate}

The numerical implementation we propose for carrying out this scheme is 
quite simple: 
\begin{enumerate}[(i)]
\item new \acp{BMR} are continuously deposited at the solar surface, at 
times, latitudes and longitudes, tilts, angular separations, magnetic 
fluxes and polarity generated through a (probabilistic) flux emergence 
algorithm based on the strength and spatial distribution of the 
deep-seated magnetic fields; 
\item the \ac{SFT} equation is solved on the solar spherical surface, 
and generates the expected cancellation, decay, transport and specific 
features typically observed in surface magnetograms 
(see Paper~I); 
\item the \ac{FTD} equation is solved in the meridional plane, using the 
evolving results of the surface simulation as a time-dependent upper 
boundary condition on the poloidal field; 
transport of this poloidal field to the base of the convection zone and 
subsequent shearing by differential rotation eventually builds up strong 
toroidal magnetic fields deep in the convection zone; 
\item the dynamo loop is closed by allowing this deep-seated magnetic 
structure to generate the emergences required in step (i).
\end{enumerate}



\subsection{Basic Ingredients} \label{s_basic}


In the depths of the solar convection zone or in the tangles of 
photospheric turbulent motions, magnetic fields are dispersed, 
transported, amplified or destroyed by small and large-scale flows. 
In the solar interior and photosphere, these processes are well-captured 
by the \ac{MHD} induction equation:
\begin{equation}
   \pd{\vec{B}}{t} = \curl (\vec{u} \cprod \vec{B} - \eta \curl \vec{B}) 
   \ , \label{eq_mhd}
\end{equation}
with $\eta$ the net magnetic diffusivity, including contributions from 
the small microscopic magnetic diffusivity $\eta_e = c^2/4\pi\sigma_e$ 
(with $\sigma_e^{-1}$ the electric resistivity of the plasma), as well 
as a dominant turbulent contribution associated with the destructive 
folding of magnetic field lines by small-scale convective fluid motions. 
We adopt here the kinematic approximation, whereby the flow $\vec{u}$ is 
considered given. 
This approximation has been shown to be appropriate in reproducing 
reasonably well the synoptic evolution of the solar surface magnetic 
field \citep[see, e.g.,][]{Wang2002a,Baumann2004}, as well as the 
overall solar dynamo properties 
\citep[see, e.g.,][and references therein]{Karak2014}. 
On spatial scales much larger than convection, two flows contribute 
to $\vec{u}$: meridional circulation $\vec u_\text{P}(r,\theta)$ and 
differential rotation $\varpi\Omega(r,\theta) \uvec{e}_\phi$. 
Both these flows can be considered axisymmetric 
($\partial/\partial\phi\equiv 0$) and steady 
($\partial/\partial t\equiv 0$) as per the kinematic approximation. 
They can be expressed in spherical polar coordinates $(r,\theta,\phi)$ as 
\begin{equation}
   \vec{u}(r,\theta) = \underbrace{\frac{R}{\rho(r)/\rho_0} \curl 
   (\Psi(r,\theta) \uvec{e}_\phi)}_{\vec{u}_\text{P}(r,\theta) = 
   u_r(r,\theta) \uvec{e}_r + u_\theta(r,\theta) \uvec{e}_\theta} + 
   \varpi \Omega(r,\theta) \uvec{e}_\phi
   \ , \label{eq_flow}
\end{equation}
where the meridional flow has been formulated in terms of a stream 
function $\Psi(r,\theta)$, thus ensuring mass conservation in a 
$\rho(r)=\rho_0\xi^m$ density profile, with $\xi(r)=(R/r)-1$, 
$m=1.5$ for an adiabatic stratification, $R$ the solar radius, 
and $\varpi=r\sin\theta$. 

\subsubsection{Meridional Circulation} \label{s_mercirc}

We opted to use a modified form of the meridional flow profile 
introduced by \citet{vanBalle1988}. 
This flow can be defined through a separable stream function of the 
form: 
\begin{subequations}
\begin{equation}
   \Psi(r,\theta) 
   = u_\theta(R,\theta)
   \frac{R}{r} \left[{-}\frac{\xi^{m+1}}{m+1} {+}
   \frac{c_1 \xi^{2m+1}}{2m+1} {-}\frac{c_2 \xi^{2m+p+1}}{2m+p+1}\right]
   \ , \label{eq_mercirca}
\end{equation}
where
\begin{alignat*}{4}
   &c_1 &&= \frac{(2m+1)(m+p)}{(m+1)p} &&\xi_b^{-m}     &&\ ,\\
   &c_2 &&= \frac{(2m+p+1)m}  {(m+1)p} &&\xi_b^{-(m+p)} &&\ ,
\end{alignat*}
and $\xi_b=(R/R_b)-1$. 
Parameters $p$ and $m$ determine the depth and concentration of the 
return flow, down to $r=R_b$. 
For the purpose of the foregoing analysis and calibration, parameters 
$p$ and $R_b$ will be treated as free parameters, while the polytropic 
index $m$ is set at $1.5$, appropriate for an adiabatic stratification. 

We deviate from the original formulation of \citet{vanBalle1988} by 
using the following latitudinal dependence, also used in Paper~I:
\begin{equation}
   u_\theta(R,\theta) = -\frac{u_0}{u_0^\ast}~ 
   \erf^q\bigl(v\sin\theta\bigr) ~\erf^n\bigl(w\cos\theta\bigr)
   \ , \label{eq_mercircb}
\end{equation}
\end{subequations}
with $u_0^\ast$ a normalization factor such that $u_0$ is the maximum 
meridional flow velocity and $q$, $n$, $v$, and $w$ parameters that 
allow to generate a very wide range of solar-like surface meridional 
flow profiles. 
The value of $n$ is fixed to $1$ as to prevent the formation of a 
$\unit{0}{\mps}$ plateau near the equator. 
We developed this flexible formulation in Paper~I to allow for the 
inclusion of various profiles used in flux transport modeling 
(e.g., \citealt{Dikpati1999,vanBalle1988,Wang2002b})
and measured on the Sun (e.g., \citealt{Ulrich2010}). 

\subsubsection{Differential Rotation} \label{s_diffrot}

Unlike meridional circulation, the solar internal differential 
rotation profile is well constrained by helioseismology.
We use here the helioseismically-calibrated solar-like
parameterization introduced in \citet{Charbonneau1999}:
\begin{subequations}
\begin{equation}
   \Omega(r,\theta) = \Omega_c + \frac{\Omega(R,\theta)-\Omega_c}{2}
   \left[ 1 + {\rm erf}\left(\frac{r-R_c}{\delta_c/2}\right) \right]
   \ , \label{eq_rotdiff}
\end{equation}
with $\Omega_c = \unit{2.724}{\micro\radps}$, $R_c=0.7R$, and 
surface rotation 
\begin{equation}
   \Omega(R,\theta) = 
   \Omega_0 \left( 1 + a_2 \cos^2\theta + a_4 \cos^4\theta \right) 
   \ , \label{eq_surfdiffrot}
\end{equation}
\end{subequations}
where $a_2=-0.1264$, $a_4=-0.1591$, 
and $\Omega_0 = \unit{2.894}{\micro\radps}$ 
(see also \citealt{Snodgrass1983}). 
The thickness $\delta_c$ of the transition region between differential 
and solid rotation, the tachocline, near the base of the convection zone, 
is kept as a free parameter. 

\subsubsection{Magnetic Diffusivity} \label{s_magdiff}

In the stably stratified core, the presumed absence of turbulence 
suggests a net diffusivity ($\eta_c$) given by Ohmic dissipation, while 
in the bulk of the convection zone, enhanced turbulent dissipation 
($\eta_t$) of the magnetic field is expected to dominate. 
The following parametric profile, given by \citet{Dikpati1999}, 
allows for a smooth transition between these two regimes: 
\begin{equation}
   \eta(r) = \eta_c + \frac{\eta_t}{2}
   \left[ 1 + {\rm erf}\left(\frac{r-R_c}{\delta_c/2}\right) \right]
   \ ,
\end{equation}
where $R_c$ takes the same value as in the preceding differential 
rotation profile.

In the surface layer, supergranular convective motions 
drive a random walk that disperses magnetic flux, and can be modeled 
as a diffusive process \citep{Leighton1964} characterized by an 
effective magnetic diffusivity of order 
$\eta_R\simeq\unit{10^{12}-10^{13}}{\cmmps}$. 
This value is used solely in the \ac{SFT} part of the model. 
The overall radial profile of $\eta(r)$ consequently includes an 
implicit step function at $r=R$. 
The exact values for $\eta_c$, $\eta_t$, and $\eta_R$, as well as 
$\delta_c$, are virtually impossible to determine from first 
principles, such that they must be treated as unknown parameters 
needing a proper calibration. 

\subsection{The Flux Transport Dynamo Equations} \label{s_dyn}

The large-scale axisymmetric magnetic field simulated in the \ac{FTD} 
component of the model can be expressed as
\begin{equation}
   \vec{B}(r,\theta,t) = \underbrace{
   \curl (A_\phi(r,\theta,t) \uvec{e}_\phi)}_
   {\vec{B}_\text{P} = B_r \uvec{e}_r + B_\theta \uvec{e}_\theta} + 
   B_\phi(r,\theta,t) \uvec{e}_\phi
   \ , \label{eq_baxi}
\end{equation}
where $\vec{B}_\text{P}$ and $B_\phi \uvec{e}_\phi$ are respectively 
the poloidal and toroidal vector components of the field. 
Inserting this decomposition for $\vec{B}$, along with 
Equation~(\ref{eq_flow}) for the flow, into the \ac{MHD} induction 
Equation~(\ref{eq_mhd}) then yields the usual two 
evolutionary equations for the scalar components $A_\phi(r,\theta,t)$ 
and $B_\phi(r,\theta,t)$:
\begin{subequations}
\label{eq_mhdaxi}
\begin{align}
   \pd{A_\phi}{t} =
   -\frac{1}{\varpi} (\vec{u}_\text{P} \cdot \grad)(\varpi A_\phi) 
   + \eta \left(\nabla^2-\frac{1}{\varpi^2}\right) A_\phi&
   \label{eq_mhdaxia}
   \ , \\ 
   \pd{B_\phi}{t} =
   -\varpi (\vec{u}_\text{P} \cdot \grad)
   \left(\frac{B_\phi}{\varpi}\right)
   + \eta\left(\nabla^2-\frac{1}{\varpi^2}\right) B_\phi&
   \nonumber \\
   - (\div\vec{u}_\text{P}) B_\phi
   + \frac{1}{\varpi}\pd{\eta}{r} \pd{(\varpi B_\phi)}{r}
   + \varpi \vec{B}_\text{P} \cdot \grad\Omega&
   \label{eq_mhdaxib}
   \ .
\end{align}
\end{subequations}
These two equations are linear in $A_\phi$ and $B_\phi$, but are coupled 
by the shearing term in Equation~(\ref{eq_mhdaxib}) which acts as a 
source for $B_\phi$ proportional to $A_\phi$. 
No such source appears explicitly in Equation~(\ref{eq_mhdaxia}). 
Here the regeneration and amplification of the poloidal field is 
supplied by a continuous input from the \ac{SFT} simulation, providing 
a time-evolving surface boundary condition for $A_\phi$ which 
effectively acts as a source.

\subsection{Surface Flux Transport} \label{s_surf}

Following earlier modeling work on surface magnetic flux evolution, 
in particular in Paper~I, we consider the magnetic field to be 
predominantly radial on global scales and we solve only the 
$r$-component of Equation~(\ref{eq_mhd}), at $r=R$. 
This leads to the usual two-dimensional linear advection--diffusion 
equation for the scalar component $B_R = B_r(R,\theta,\phi,t)$,
\begin{align}
   \pd{B_R}{t} =
   &- \frac{1}{R \sin\theta} 
   \pd{}{\theta}\big[ u_\theta(R,\theta) B_R \sin\theta \big] - 
   \Omega(R,\theta) \pd{B_R}{\phi} \nonumber \\
   &+ \frac{\eta_R}{R^2} \left[ \frac{1}{\sin\theta} 
   \pd{}{\theta}\left(\sin\theta\pd{B_R}{\theta}\right) + 
   \frac{1}{\sin^2\theta} \pdd{B_R}{\phi} \right] \nonumber \\
   &- \frac{B_R}{\tau_R} + S_\text{BMR}(\theta,\phi,t) \ ,
   \label{eq_surftrans}
\end{align}
to which two supplementary terms have been added: 
a source term $S_\text{BMR}(\theta,\phi,t)$ to account for the 
discrete emergence of new surface flux in the form of \acp{BMR}, 
and a linear sink term $-B_R/\tau_R$ to allow for an exponential decay 
of the surface field with time. 
\citet{Schrijver2002} originally found such a decay on a timescale 
of $\unit{5-10}{\years}$ to be necessary to preclude secular drift and 
ensure polarity reversal of the polar caps when modeling surface flux 
evolution over many successive cycles of differing amplitudes. 
This was subsequently justified physically by \citet{Baumann2006} as the 
effect of a vertical turbulent diffusion, or equivalently a convective 
submergence, on the decay of the dominant dipole mode, two physical 
mechanisms that cannot be directly included in the \ac{SFT} model. 
We included this term in Paper~I but did not find it to be 
required for the \ac{SFT} results to match the synoptic magnetogram of 
cycle 21. 
We test it again here, with $\tau_R$ treated as a free parameter.

\subsection{Numerical Solution and Coupling} \label{s_grid}

The \ac{FTD} equations~(\ref{eq_mhdaxi}) and 
\ac{SFT} equation~(\ref{eq_surftrans}) are solved concurrently, each on 
a separate two-dimensional computational grid on which spatial 
discretization is carried out via the Galerkin finite element method, 
and implicit temporal discretization through the one-step $\Theta$-method 
(see, e.g., \citealt{Burnett1987}).

The \ac{SFT} simulation is solved over a regular Cartesian grid in 
$(\theta,\phi)$ representing the whole solar surface, with longitudinal 
periodicity enforced through a padding of ghost cells.
Rigorous flux conservation is also required since only a small fraction 
of the emerging magnetic flux ultimately builds up the axial dipole 
observed at sunspot minima.
We minimize numerical discretization errors by adopting double precision 
arithmetics, a $256\times128$ longitude--latitude grid, and $8000$ 
time steps for the eight-cycle runs that will be analyzed in \S~3 
(for more details on numerical errors see Paper~I, 
\S~2.4 and discussion therein).

The \ac{FTD} simulation is solved simultaneously over a regular 
$96\times128$ Cartesian grid in ($r$,$\theta$), from pole to pole 
and $0.5 \leq r/R \leq 3.0$. 
Below $r=0.5R$, the radiative core is considered perfectly conductive 
and the $A_\phi=B_\phi=0$ boundary condition is applied. 
For $r > R$, the absence of flows and electrical currents imposes 
$B_\phi=0$.
The spherical geometry finally constrains $A_\phi=B_\phi=0$ at the 
poles. 
The overall scheme is similar to that described in 
\citet{Charbonneau2005}. 

With such spatial resolutions and typical time steps of $\simeq 4$ and 
$\simeq40$ days respectively in the SFT and FTD simulations, the former 
dominates the computational workload by a factor of $\simeq20$.

\subsubsection{From \ac{SFT} to \ac{FTD}}
The surface ($r=R$) boundary condition on $A_\phi$ is updated at every
\ac{FTD} time step, via the longitudinal averaging of the \ac{SFT} 
solution $\left(\langle B_R \rangle^\phi (\theta,t)\right)$ and 
integration of the resulting latitudinal function:
\begin{equation}
   A_\phi(R,\theta,t) = A_\phi^0 + \frac{R}{\sin\theta} 
   \int \langle B_R \rangle^\phi (\theta,t) \sin\theta d\theta  \ ,
\end{equation}
where $A_\phi^0$ is set to zero at the poles. 
This provides the coupling from the \ac{SFT} toward the \ac{FTD} model.

Such coupling assumes that physical processes responsible for surface
magnetic flux evolution occur only inside the single \ac{FTD} grid layer 
located at $r=R$, which is of thickness $\simeq\unit{3.7}{\mega\meter}$ 
for our working spatial mesh.

\subsubsection{From \ac{FTD} to \ac{SFT}: Emergence Function} \label{s_emerg}

The coupling from the \ac{FTD} toward the \ac{SFT} is the emergence of 
\acp{BMR}. 
In view of the considerable complexity of the various processes 
involved in the formation, destabilization, buoyant rise, and emergence 
of deep-seated magnetic flux tubes 
(see, e.g., \citealt{Weber2011} and review by \citealt{Fan2009}), 
we opted here to input emerging \acp{BMR} directly into the \ac{SFT} 
component of the model, based on a semi-empirical emergence function 
giving, as a function of the strength of the internal magnetic field, 
the {\it probability} that the emergence of a \ac{BMR} will occur. 

Calculations of the destabilization and buoyant rise of magnetic 
flux tubes carried out in the thin-tube approximation do offer some 
useful guidance.
From the stability diagrams obtained by \citet{Schussler1994} and 
\citet{FerrizMas1994}, one can infer the depth, latitude and magnetic 
amplitudes at which toroidal flux tubes are expected to destabilize. 
According to their results, and depending on the level of 
subadiabaticity in the outer reaches of the radiative core, instability 
growth rates near $r/R\simeq0.7$ remain 
approximately constant, or show a smooth increase with latitude, from 
the equator up to $\simeq\unit{70}{\degree}$, and then fall of rapidly 
to zero over a latitudinal width of $\simeq\unit{5}{\degree}$. 
A lower threshold of order $\unit{10^4-10^5}{\gauss}$ is also required, 
on the amplitude of the magnetic field inside concentrated flux tubes. 
A crucial missing link is the degree of magnetic field amplification 
taking place during the formation of these toroidal flux tubes from
the dynamo-generated large-scale magnetic field. 
Accordingly, we define this lower limit as $B^\ast\in[10^1,10^4]$ 
(with units that depend on the exact parameterization of 
Equation~\ref{eq_fitfunc} below), 
and treat it as another free parameter to be calibrated. 
Modeling also shows that a certain level of twist is required for the 
tube to maintain its coherence during the rise through the convective 
envelope \citep{Fan2009}.
Accordingly, we introduce the quantity 
$\abs{B_\text{mix}} = \abs{B_\phi}^b \abs{A_\phi}^a$,  
evaluated at depth $r^\ast/R \in [0.60,0.80]$ and 
with exponents in the ranges $b\in[0.5,3.0]$ and $a\in[0.0,2.0]$, 
and use it to build the following quasi-normalized emergence function:
\begin{align}
   \abs{F_B(\theta,t)} 
   &= \frac{1}{4}\left(1+\erf\left(\frac{\abs{B_\text{mix}}-B^\ast}
   {\delta B^\ast}\right)\right)
   \abs{\frac{B_\text{mix}}{\text{max}\abs{B_\text{mix}}}}^c
   \nonumber \\
   &\times \left(\bigl(1-\mu_\ell\bigr)\frac{\abs{\ell}}{90}+\mu_\ell\right)
   \left(1-\erf\left(\frac{\abs{\ell}-\ell^\ast}
   {\delta\ell^\ast}\right)\right)
   \ . \label{eq_fitfunc}
\end{align}
The first part of Equation~(\ref{eq_fitfunc}) sets a lower threshold on
$B_\text{mix}$ above which emergences can take place,
as well as a possible saturation ($c\rightarrow0$) or linear growth 
($c\rightarrow1$) of the probability above $B^\ast$. 
The transition scale $\delta B^\ast$ is set to some fraction of 
$B^\ast$ (see \S~\ref{s_calib}). 
The second part accounts for the latitudinal dependence of the 
instability's growth rate, which we assume to increase linearly from 
$\mu_\ell \in [0,1]$ at the equator to $1$ near latitude 
$\ell^\ast\in[\unit{65}{\degree},\unit{90}{\degree}]$, followed by a 
quick drop to zero in $\delta\ell^\ast=3^\circ$ 
(cf. Figures~1 and 2 of \citealt{FerrizMas1994}). 
The sign of $F_B(\theta,t)$ is given by the sign of the input $B_\phi$.

The emergence process is made inherently non-deterministic with the 
following sources of stochasticity:
\newcounter{saveenum}
\begin{enumerate}[(i)]
\item at every \ac{SFT} time step, the number $N(t)$ of new \acp{BMR} 
to emerge is extracted from a uniform random distribution, proportional 
to the sum $\sum_\theta F_B(\theta,t)$ at the corresponding \ac{FTD} 
time step; 
\item the probability of emergence of a \ac{BMR} at a given latitude 
is made proportional to $F_B(\theta,t)$.
\setcounter{saveenum}{\value{enumi}}
\end{enumerate}
Also, independently from the distribution of $F_B(\theta,t)$, and as 
determined in our analysis of \cite{Wang1989-0}'s database entries
(see Appendix A of Paper~I):
\begin{enumerate}[(i)]
\setcounter{enumi}{\value{saveenum}}
\item emergence longitudes are assumed to be random;
\item magnetic fluxes $\Phi$ are extracted from a log-normal 
distribution centered at $\log\Phi_0=\unit{21.3}{(\log\maxwell)}$ with 
standard deviation $\sigma_{\log\Phi}=\unit{0.5}{(\log\maxwell)}$ 
(Paper~I, Equation~(13)),
independently of cycle phase and amplitude 
(following \citealt{Bogdan1988}); 
\item magnetic bipole separations $\delta$ follow a power law with flux, 
with a gaussian dispersion about it (Paper~I, Equation~(15)); 
\item magnetic bipole tilts $\alpha$ relative to the equatorial 
direction follow a linear increase with latitude (Joy's law) and a 
gaussian spread with standard deviation decreasing exponentially with 
$\log\Phi$ (Paper~I, Equations~(16{\it a}) and (16{\it b})).
\end{enumerate}

The input of \acp{BMR} in the \ac{SFT} simulation enters the source term
\begin{subequations}
\begin{equation}
   S_\text{BMR}(\theta,\phi,t) = \sum_{i=1}^{N(t)} B_i(\theta,\phi)
   \delta_D(t-t_i) \ ,
   \label{eq_sbmr}
\end{equation}
with $\delta_D$ the Dirac delta. 
Each new \acp{BMR} is placed at its given position $(\theta_i,\phi_i)$ 
and time $t_i$, with a gaussian distribution for each pole:
\begin{equation}
   B_i(\theta,\phi) = 
   \underbrace{ B_{i0}e^{-\delta_{i+}^2/2\sigma^2}}_{B_{i+}(\theta,\phi)} 
   +
   \underbrace{-B_{i0}e^{-\delta_{i-}^2/2\sigma^2}}_{B_{i-}(\theta,\phi)} 
   \ ,
   \label{eq_bbmr}
\end{equation}
\end{subequations}
where $\delta_{i+}$ and $\delta_{i-}$ are the heliocentric angular 
distances from  the centres $(\theta_{i+},\phi_{i+})$ and 
$(\theta_{i-},\phi_{i-})$ of the two poles, respectively, and 
$\sigma=\unit{4}{\degree}$ the width of the gaussians.

The preceding steps dictate the relative probability of given emergences 
to occur, but the actual number $N(t)$ of \acp{BMR} to emerge every time 
step remains adjustable. 
We introduce a proportionality factor $K$ between the emergence function 
$F_B(\theta,t)$ and the actual emerged butterfly diagram, so that
$N(t) = K\cdot\sum_\theta F_B(\theta,t)$. 
Therefore, $K$ effectively acts as a dynamo number in the model.
Here however, the fact that the poloidal source term depends on a 
{\it number} of emergences $N(t)$, rather than being directly 
proportional to the underlying toroidal flux, means that the 
relationship is not formally linear. 
Nonetheless, as described in \S~\ref{s_bldynlin}, the model appears to 
behave linearly when averaged over many different stochastic 
realizations of emergences. 
Stochastic aspects notwithstanding, $K$ may thus be considered a dynamo 
number in a statistical sense, as it sets the mean growth rate in the 
linear regime.
This dynamo number is akin to that encountered in the classical
mean-field framework, where it is defined as the dimensionless product 
of the strength of differential rotation and turbulent electromotive 
force over magnetic dissipation. 
Moreover, as demonstrated by the dynamo solutions to be discussed 
presently, the value of $K$ also sets the absolute mean amplitude 
of the dynamo, together with the tilt-quenching mechanism introduced 
in \S~\ref{s_bldynqch}.

As a result, for the reference dynamo solution presented in 
\S~\ref{s_bldynqch}, with the working spatial mesh and time stepping 
described above and after adjustment of $K$ to obtain stable, 
solar-like solutions, the value of $N(t)$ varies from 
$5-15$ per \ac{SFT} time step ($50-150$ per month) near cycle maxima 
down to $0-1$ per \ac{SFT} time step ($0-10$ per month) at cycle minima. 

Meanwhile, the exact distribution of these newly emerged \acp{BMR}, 
i.e. the shape of $F_B(\theta,t)$, is mostly critical if one strives 
to match the observed butterfly diagram. 
The next logical step is now to carry out a calibration of all 
parameters describing the full model, using observed emergences as a 
constraint, as detailed in the following section.

\section{Model Calibration} \label{s_calib}

The various physical components of the coupled \ac{SFT}--\ac{FTD} 
model introduced in the preceding section jointly involve a large number 
of numerical parameters; $27$ to be precise. 
Nine of these can be fixed confidently either through observations 
or theoretical considerations. 
Five ($R_c$, $\Omega_0$, $\Omega_c$, $a_2$, and $a_4$) are the 
numerical parameters defining the differential rotation profile 
(see \S~\ref{s_diffrot}), another ($m$) is the polytropic index 
characterizing the stratification within the convection zone, and yet 
another ($n$) is used to formulate a flexible surface meridional flow 
profile but set to $1$ to reflect solar observations 
(see \S~\ref{s_mercirc}). 
The last two parameters to be held fixed, $\delta B^\ast$ and 
$\delta\ell^\ast$, control the shapes of the latitudinal and magnetic 
masking used in the emergence function (see \S~\ref{s_emerg}); 
experimenting with the model reveals that within reasonably wide 
ranges, the exact values chosen for these parameters have little impact 
on the global dynamo behavior. 
Consequently, they are fixed at values $0.1B^\ast$ and 
$\unit{3}{\degree}$ respectively.

This leaves $18$ adjustable parameters, which are listed in 
Table~\ref{t_param}. 
Eleven pertain to the linear terms in the model, including the shape of 
the meridional flow, magnetic diffusivity and surface sink ($\delta_c$, 
$R_b$, $u_0$, $p$, $q$, $v$, $w$, $\eta_c$, $\eta_t$, $\eta_R$, and $
\tau_R$), 
and the remaining seven ($r^\ast$, $b$, $a$, $c$, $B^\ast$, $\ell^\ast$, 
and $\mu_\ell$) to the form of the nonlinear emergence function 
(Equation~(\ref{eq_fitfunc})).

\begin{deluxetable}{lrrr}
\tablecaption{Parameter values\label{t_param}}
\tablehead{
\colhead{Parameters}&\colhead{\tblmk{b}Reference}& \colhead{Tested}                       & \colhead{\tblmk{g}Optimal}                                  \\*[0.0ex]
                        & \colhead{Values}       & \colhead{Intervals}                    & \colhead{Values}                                            \\*[1.0ex]
                        & \colhead{($C=0.42$)}   & \colhead{}                             & \colhead{($C\in[0.92,0.94]$)}                                         }
\startdata
$r^\ast/R$             &            $0.705\quad$ &          $[0.60\phm{^0},0.80\phm{^0}]$ & $\mathbf{\phm{0}0.68}\pm\substack{0.04\phm{}\\0.03\phm{}}$~~ \\*[1.0ex]
$b$                    &              $1.0\quad$ &          $[0.5\phm{0^0},3.0\phm{0^0}]$ & $\mathbf{\phm{00}1.5}\pm\substack{1.5\phm{0}\\0.5\phm{0}}$~~ \\*[1.0ex]
$a$                    &              $0.0\quad$ &          $[0.0\phm{0^0},2.0\phm{0^0}]$ & $\mathbf{\phm{00}0.0}\pm\substack{0.8\phm{0}\\0.0\phm{0}}$~~ \\*[1.0ex]
$c$                    &              $1.0\quad$ &          $[0.0\phm{0^0},1.0\phm{0^0}]$ & $\mathbf{\phm{00}1.0}\pm\substack{0.0\phm{0}\\0.6\phm{0}}$~~ \\*[1.0ex]
$B^\ast~\tblmk{a}$     &  \tblmk{c}$~~10^2\quad$ &          $[10^1\phm{.0},10^4\phm{.0}]$ &        $\mathbf{10^2}\pm\substack{0\phm{.00}\\10^2\phm{}}$~~~\\*[1.0ex]
$\ell^\ast$            &               $45\quad$ &          $[64\phm{0.^0},90\phm{0.^0}]$ & $\mathbf{\phm{.00}70}\pm\substack{9\phm{.00}\\6\phm{.00}}$~~ \\*[1.0ex]
$\mu_\ell$             &              $0.0\quad$ &          $[0.0\phm{0^0},1.0\phm{0^0}]$ & $\mathbf{\phm{00}0.5}\pm\substack{0.5\phm{0}\\0.5\phm{0}}$~~ \\*[1.0ex]
$\delta_c/R$           &             $0.05\quad$ &          $[0.04\phm{^0},0.10\phm{^0}]$ & $\mathbf{\phm{0}0.05}\pm\substack{0.02\phm{}\\0.01\phm{}}$~~ \\*[1.0ex]
$R_b/R$                &             $0.69\quad$ &          $[0.60\phm{^0},0.70\phm{^0}]$ & $\mathbf{\phm{0}0.60}\pm\substack{0.02\phm{}\\0.00\phm{}}$~~ \\*[1.0ex]
$u_0$/\mps             &               $18\quad$ &\tblmk{e}~$[8\phm{0.0^0},18\phm{0.^0}]$ & $\mathbf{\phm{.00}17}\pm\substack{1\phm{.00}\\8\phm{.00}}$~~ \\*[1.0ex]
$\log p$               &              $2.0\quad$ &          $[{-}1.0\,,2.0\phm{0^0}]$     &      $\mathbf{{-}0.7}\pm\substack{1.2\phm{0}\\0.2\phm{0}}$~~ \\*[1.0ex]
$q$                    &              $2.5\quad$ &\tblmk{f}~$[2^0\phm{0.0},2^5\phm{0.0}]$ & $\mathbf{\phm{.000}1}\pm\substack{31\phm{.0}\\0\phm{.00}}$~~ \\*[1.0ex]
$v$                    &              $1.0\quad$ &\tblmk{f}~$[2^0\phm{0.0},2^3\phm{0.0}]$ & $\mathbf{\phm{.000}7}\pm\substack{1\phm{.00}\\5\phm{.00}}$~~ \\*[1.0ex]
$w$                    &              $3.5\quad$ &\tblmk{f}~$[2^0\phm{0.0},2^5\phm{0.0}]$ & $\mathbf{\phm{.000}1}\pm\substack{1\phm{.00}\\0\phm{.00}}$~~ \\*[1.0ex]
$\log(\eta_c$/\cmmps)  &                $9\quad$ &          $[7\phm{0.0^0},11\phm{0.^0}]$ & $\mathbf{\phm{00}8.0}\pm\substack{2.4\phm{0}\\1.0\phm{0}}$~~ \\*[1.0ex]
$\log(\eta_t$/\cmmps)  &             $10.7\quad$ &          $[11.0\phm{^0},13.0\phm{^0}]$ & $\mathbf{\phm{0}12.0}\pm\substack{0.2\phm{0}\\0.4\phm{0}}$~~ \\*[1.0ex]
$\log(\eta_R$/\cmmps)  &            $12.48\quad$ &\tblmk{e}~$[12.38\phm{},12.82\phm{}]$   & $\mathbf{\phm{}12.78}\pm\substack{0.04\phm{}\\0.40\phm{}}$~~ \\*[1.0ex]
$\tau_R$/years         &  \tblmk{d}$~~~~32\quad$ &\tblmk{e}~$[7\phm{0.0^0},32\phm{0.^0}]$ & $\mathbf{\phm{.00}10}\pm\substack{\infty\phm{0}\\3\phm{.0}}$~~~
\enddata
\tablenotetext{}{\textbf{Notes.}}
\tablenotetext{a}{
The units of $B^\ast$ depend on the values of exponents $b$ and $a$, 
since they must be the same than the units of 
$\abs{B_\text{mix}} = \abs{B_\phi}^b \abs{A_\phi}^a$ to ensure coherence 
in Equation~(\ref{eq_fitfunc}).}
\tablenotetext{b}{
Reference values as to approximate velocity and diffusivity profiles 
and emergence algorithm used by MD2014, leading to the 
solution shown in Figure~\ref{f_W21x8}(a).}
\tablenotetext{c}{
Threshold value $B^\ast$ unavailable from MD2014.}
\tablenotetext{d}{
$\tau_R \gtrsim \unit{32}{\years}$ is similar to removing term 
$-B_R/\tau_R$ in Equation~(\ref{eq_surftrans}).}
\tablenotetext{e}{
As determined in Paper~I, where the initial interval were 
$u_0\in\unit{[5,30]}{\mps}$, $\eta_R\in\unit{[10^2,10^4]}{\kmmps}$, and 
$\tau_R\in\unit{[2^1,2^5]}{\years}$.
The linear correlation between $u_0$ and $\eta_R$ obtained from
the surface analysis should still be considered in conjunction
with the final results given in the rightmost column.}
\tablenotetext{f}{
As opposed to the optimal intervals obtained in Paper~I, where 
$w{=}8\pm\substack{24\\4}$, $v{=}2.0\pm\substack{1.5\\1.0}$, and 
$q{=}{\left(2.8\pm\substack{2.0\\1.1}\right)}\cdot{2^{1.25(\log_2v)^2}}$.}
\tablenotetext{g}{
Solutions for the first seven parameters ($r^\ast$, $b$, $a$, $c$, 
$B^\ast$, $\ell^\ast$, and $\mu_\ell$) result from the full 
W21$\times$8-18 optimization.
Solutions for the remaining eleven parameters ($\delta_c$, $R_b$, $u_0$, 
$p$, $q$, $v$, $w$, $\eta_c$, $\eta_t$, $\eta_R$, and $\tau_R$) result 
from the subsequent W21$\times$8-11 optimization.
``Optimal values'' listed in bold font correspond to one chosen optimal 
solution (see Figures~\ref{f_W21x8}(e) and (f)) among the acceptable 
solutions bounded by the given error bars. 
Other combinations of parameters allowed by the error bars should still 
be used with care, considering the shape of the parameter-space 
landscape inside the optimal region and in particular the correlations 
described at the end of \S~\ref{s_optsol}.
}
\end{deluxetable}

\subsection{Validation with the MD2014 Model}

The large number of model parameters listed in Table~\ref{t_param} 
results from the very general forms adopted for many model ingredients, 
notably the meridional flow profile and emergence function. 
This gives the model great flexibility, in that it includes as a subset 
a number of published models. 
As an example and a form of validation exercise, we now reproduce a 
dynamo solution resembling that presented in MD2014. 

Since MD2014's model includes a full two-dimensional 
representation of the solar surface and an emergence algorithm similar 
to ours, direct contact is allowed between specific features of the two 
models despite significant differences in algorithmic implementation 
and numerical procedures. 
Their (single-cell) meridional circulation profile 
(described in \citealt{Dikpati2011}) and magnetic diffusivity profile 
(described in \citealt{Dikpati2007}) may be closely approached by ours, 
through the parameter values listed in the first column of 
Table~\ref{t_param}. 
Similarly, their emergence function is comparable to the one we describe 
in \S~\ref{s_emerg}, with a latitudinal masking approximated by 
parameters $\mu_\ell=0$ and $\ell^\ast=\unit{45}{\degree}$ (a 
low-latitude cutoff conducive to the production of a solar-like 
butterfly diagram but hard to justify from the point of view of 
stability of thin flux tubes) and applied only to the $B_\phi$ component   
evaluated near depth $r^\ast/R=0.705$.
The magnetic masking includes a lower threshold $B^\ast$ of unspecified
value and apparently no upper saturation threshold (parameter $c=1$). 
The detailed parametrization of individual emerging \acp{BMR} 
nonetheless differs significantly from ours, in a generally more 
deterministic manner. 
The latitude of emergence is directly associated with the location of 
peak toroidal field, as compare to the probabilistic approach we use. 
The tilt, separation, size and flux of the spot pair are
mainly determined by the value of $B_\phi$ and the latitude of 
emergence, and so are deterministic rather than stochastic.

\begin{figure*}
   \plotone{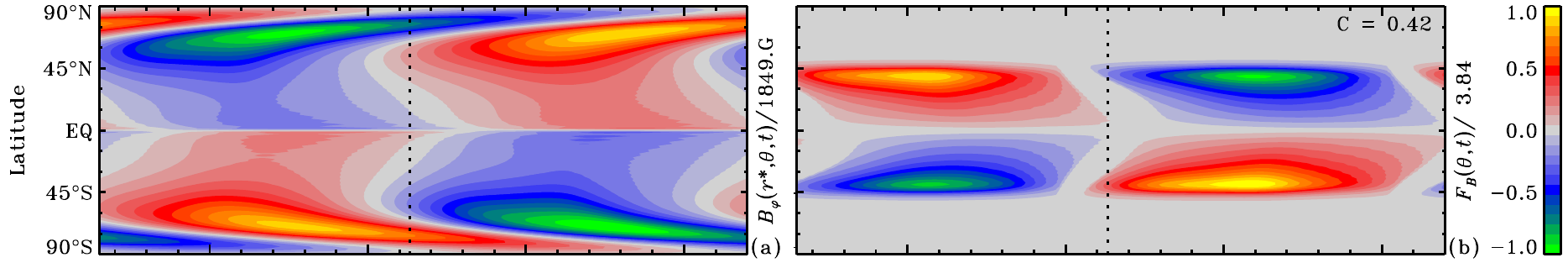}
   \plotone{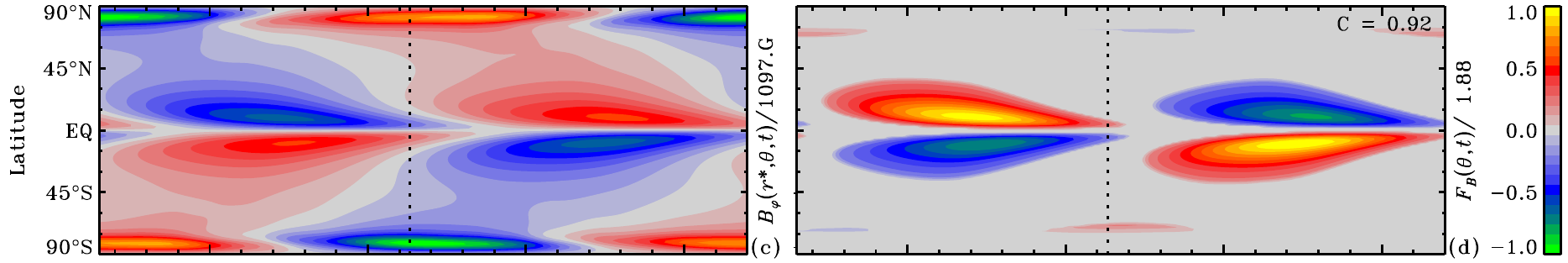}
   \plotone{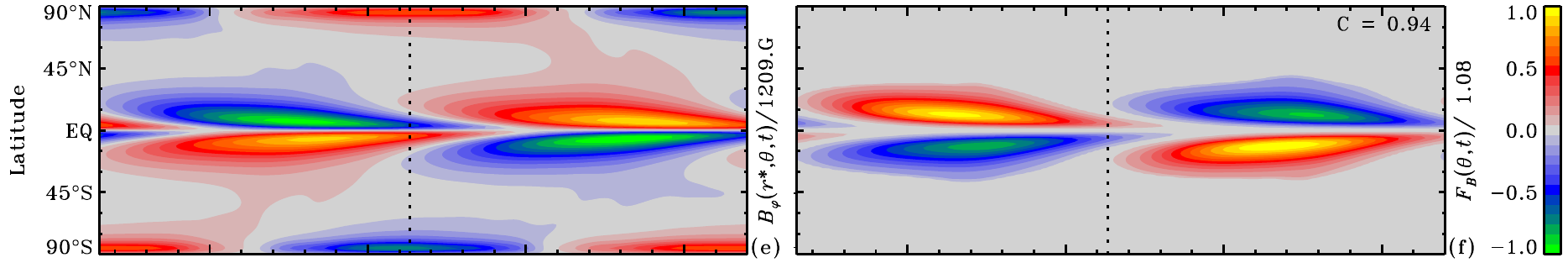}
   \plotone{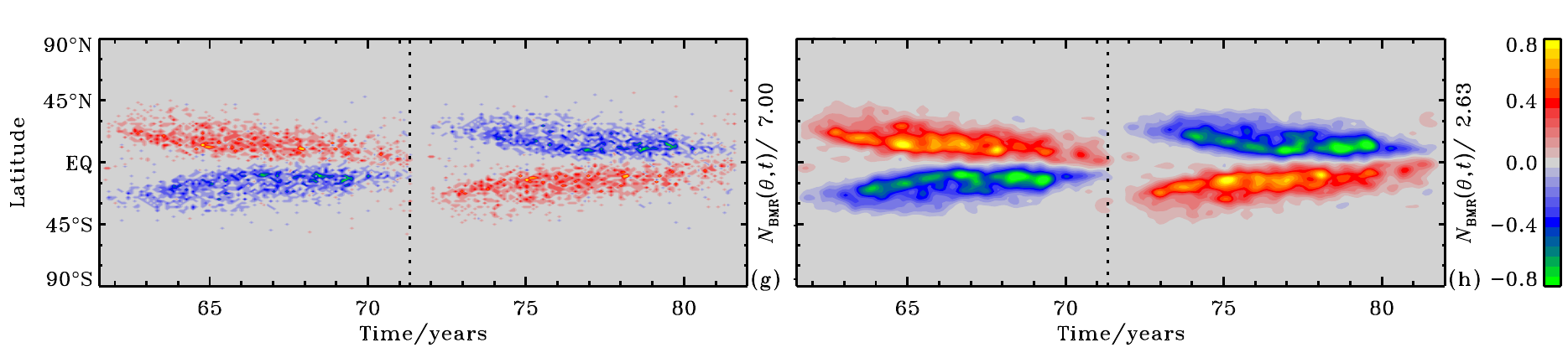}
   \caption{Left: time--latitude contour plots of the toroidal magnetic 
   field component $B_\phi(r^\ast,\theta,t)$, at $r^\ast/R=0.68$, for 
   (a) a two-cycle reference solution approaching that by MD2014, 
   (c) an example of an acceptable solution with $C=0.92$, 
   and (e) an optimal solution ($C_\text{max}=0.94$).
   (g) Raw density plot of observed \acp{BMR}, extracted from 
   \cite{Wang1989-0}'s database, where all emergences in a given 
   hemisphere and cycle have been attributed the same polarity. 
   Right: (b),(d), and (f) time-latitude contour plots of the emergence 
   function $F_B(\theta,t)$ associated with each of the solutions 
   presented at the left, with their respective fitness factor $C$. 
   (h) Smoothed version of the density plot presented at the left. 
   All diagrams show the last quarter of simulations W21$\times$8 
   (last two repetitions of cycle 21), which was used for optimization. 
   Time, given in years, starts at the beginning of the eight-cycles 
   runs. 
   Vertical dotted lines indicate the times of activity minima.}
   \label{f_W21x8}
\end{figure*}

In order to minimize the differences associated with stochastic 
realizations of our emergence procedure, we limit this exercise to the 
input of observed emergences. 
Following Paper~I, we use the comprehensive database of over 3000 
\acp{BMR} gathered by \cite{Wang1989-0} for cycle 21. 
By feeding these data into Equations~(\ref{eq_sbmr}) and 
(\ref{eq_bbmr}), the $2\times2$D simulation is indirectly forced to 
run in a cycle-21-like mode. 
The remaining model parameters are set to mimic MD2014's 
model (first column of Table~\ref{t_param}). 
We obtain the two-cycles solution presented in Figure~\ref{f_W21x8}(a), 
for the synoptic evolution of $B_\phi$ at the base of the convection 
zone. 
This solutions resembles MD2014's result in that it presents 
a strong mid-high-latitude poleward branch. Our low-latitude equatorial 
branch is however much weaker.
Applying the appropriate latitudinal and magnetic mask from
MD2014, we obtain the emergence function, or equivalently 
the probabilistic distribution of emergences, presented in 
Figure~\ref{f_W21x8}(b). 
This resembles the pattern of emergence produced in MD2014, 
with surface emergences strongly localized around $\pm 40^\circ$ 
latitude, with a hint of equatorward propagation (see their Figure~2a, 
keeping in mind that the slanted thick poleward streaks going from mid 
to high latitudes on this time--latitude plot reflect post-emergence 
surface flux transport, not emergence per se).

\subsection{Numerical Optimization} \label{s_opt}

We now seek to select model parameter values so as to obtain a 
solar-like dynamo solution. 
This defines a numerical optimization task which consists in optimizing 
the $18$ parameters listed in Table~\ref{t_param} to yield the closest 
possible fit to solar observations.

The first choice to be made is the goodness-of-fit measure to be used 
to drive such optimization. 
We opted to use a single fitness measure, namely the value of the 
linear correlation coefficient $C$ between the synoptic distribution of 
synthetic and observed emergences of \acp{BMR}. 
This presupposes that the magnetic flux tubes producing \acp{BMR} 
upon emergence through the photosphere rise radially through the 
convection zone, on a timescale very much shorter than the cycle 
period. 
Models based on the thin flux tube approximation support this idea, at 
least for the more strongly magnetized flux tube presumably producing 
the larger \acp{BMR} 
(see, e.g., \citealt{Fan2009}, and references therein).

Next we must select a suitable observational dataset against which to 
optimize the model. 
As for the preceding validation exercise, we use \cite{Wang1989-0}'s 
\acp{BMR} database for cycle 21. 
In order to minimize any influence of the initial condition (solar 
minimum-like dipolar configuration, as introduced in Paper~I), we 
generate a sequence of eight replicates of the cycle 21 database 
(hereafter W21$\times$8), by sequentially inverting the latitudes of 
emergence from one replication to the next, and use the output 
corresponding to the last two cycles to compute the correlation 
coefficient.

\subsection{\ac{GA}: PIKAIA} \label{s_pik}

We perform the numerical optimization of $C$ using the \ac{GA}-based 
optimizer PIKAIA 1.2\footnote{
\url{http://www.hao.ucar.edu/modeling/pikaia/pikaia.php} (March 2015)}
\citep{Charbonneau1995,Charbonneau2002b}.
\acp{GA} allow for an efficient and adaptive exploration of
the parameter space, and are thus quite robust at handling global 
optimization problems.
As described in Paper~I, they also allow for a quasi-Monte 
Carlo sampling of the parameter space about the current optimum 
solution, thus helping to construct error estimates on optimal 
parameter values.
In the present context PIKAIA is operating in a 18-dimensional
parameter space (viz. Table~\ref{t_param}), with the fitness measure 
given by the correlation $C$.
Calculating the fitness of a single trial solution (18-parameter 
vector) implies running the \ac{SFT} and \ac{FTD} simulations in 
parallel, with appropriate coupling through the surface boundary 
condition, and finally evaluating $C$. 
For our working spatial mesh and time stepping this requires about 
twenty minutes on a single-core modern CPU. 
For a typical optimization run of $500$ generations with $96$ 
trial solutions per generation, this adds up to $667$ core-days, but 
the fitness calculation being almost trivial to parallelize across the 
population, the wall-clock time can be brought down to a few days.

\subsection{Choosing Parameter Ranges}

PIKAIA is designed to carry out optimization in a bounded parameter
space.
The intervals explored for each parameter (second column of 
Table~\ref{t_param}) are chosen to be physically meaningful and 
computationally stable.
In particular, parameters $u_0$, $\eta_R$, and $\tau_R$ are restricted 
to the intervals found in Paper~I to better reproduce surface synoptic 
magnetograms. 
Parameters $q$, $v$, and $w$, however, are left free to vary in their 
original intervals despite the preceding calibration, to allow full 
exploration of the domain. 
Diffusivity values $\eta_c$ and $\eta_t$ and profile parameters 
$\delta_c$, $R_b$, and $p$ are given broad intervals but still within 
limits inferred by theoretical considerations and numerical experiments. 
Masking parameters are allowed to vary within ranges inferred from 
calculated stability diagrams, as described in \S~\ref{s_emerg}.

\subsection{Optimal Solution for Cycle 21} \label{s_optsol}

\begin{figure}
   \plotone{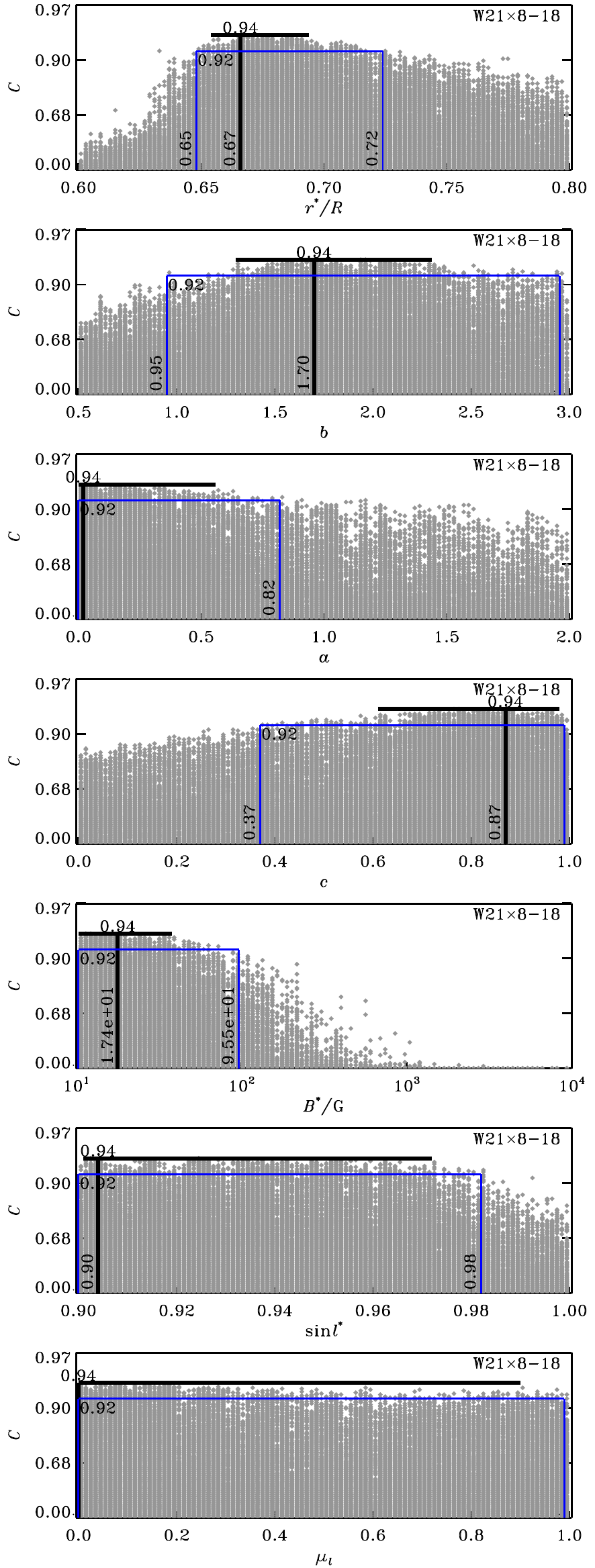}
   \caption{Distribution of the fitness $C$ (vertical inverse log 
   scale) as a function of each of the seven ``emergence'' parameters 
   ($r^\ast$, $b$, $a$, $c$, $B^\ast$, $\ell^\ast$, and $\mu_\ell$). 
   Each gray dot indicates the parameter-space position of one of the 
   $192000$ solutions obtained from four independent W21$\times$8-18 
   optimizations. 
   The remaining eleven parameters are not shown here since 
   their final analysis is based instead on the W21$\times$8-11 
   optimization (see Figure~\ref{f_fitland}). 
   On each plot, the thick horizontal line indicates the interval where
   $C\geq0.935$, and the thick vertical line the parameter value where
   true maximum fitness $C=0.94$ is reached. 
   Thin vertical blue lines delimit the parameter values where fitness
   reaches $C=0.92$, such that any solution above the horizontal blue 
   line is considered acceptable.}
   \label{f_fitland0}
\end{figure}
\begin{figure}
   \plotone{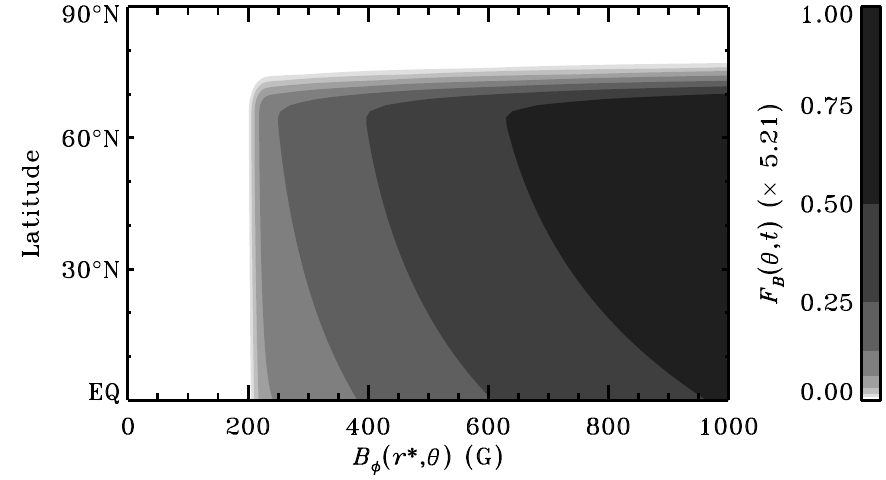}
   \caption{``Stability diagram'' used as a mask on the toroidal 
   magnetic field component $B_\phi(r^\ast,\theta,t)$ shown in 
   Figure~\ref{f_W21x8}(e), to produce the emergence function 
   $F_B(\theta,t)$ shown in Figure~\ref{f_W21x8}(f). 
   This corresponds to Equation~(\ref{eq_fitfunc}) with
   $\delta B^\ast=10^{-1}B^\ast$, and parameters $b$, 
   $a$, $c$, $B^\ast$, $\ell^\ast$, and $\mu_\ell$ set to their final 
   values, as listed in the rightmost column of Table~\ref{t_param}.
   }
   \label{f_stab}
\end{figure}
\begin{figure*}
   \plottwo{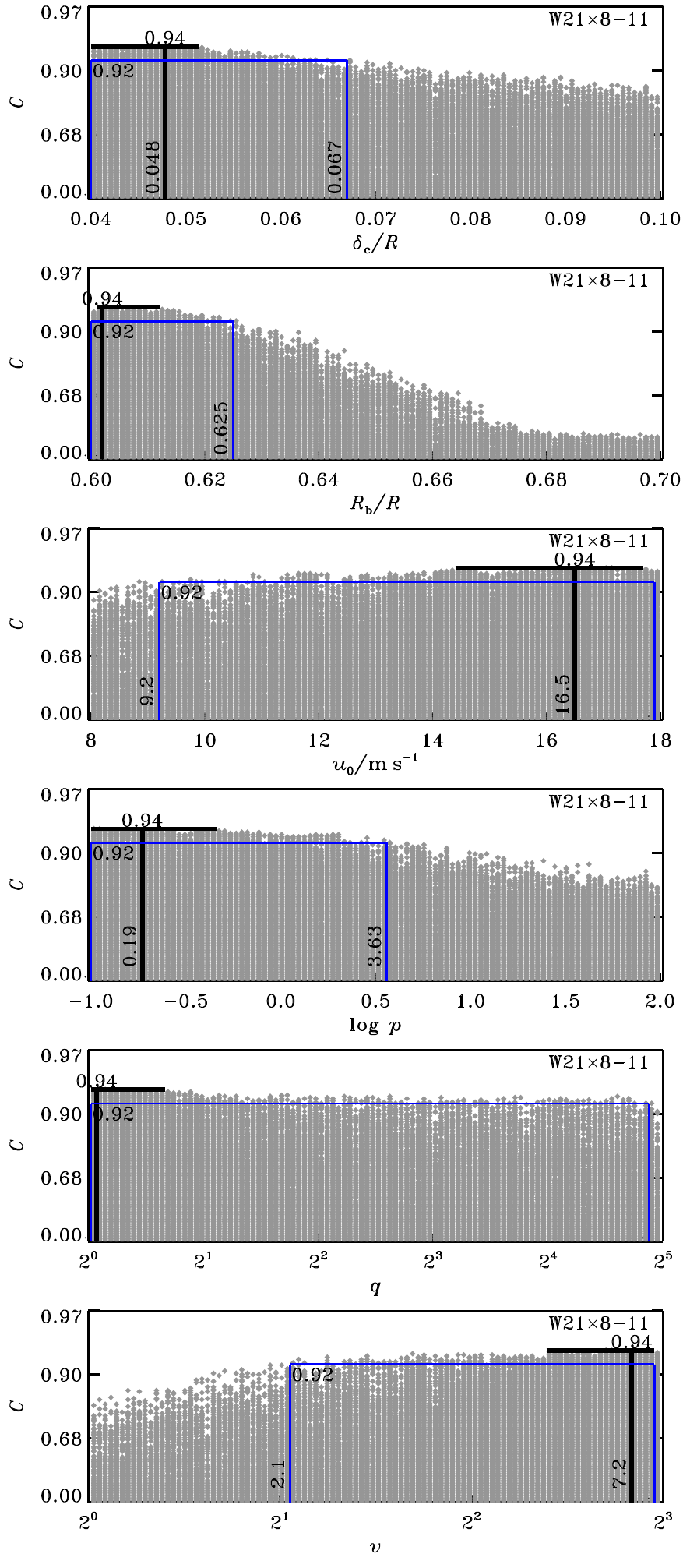}
   {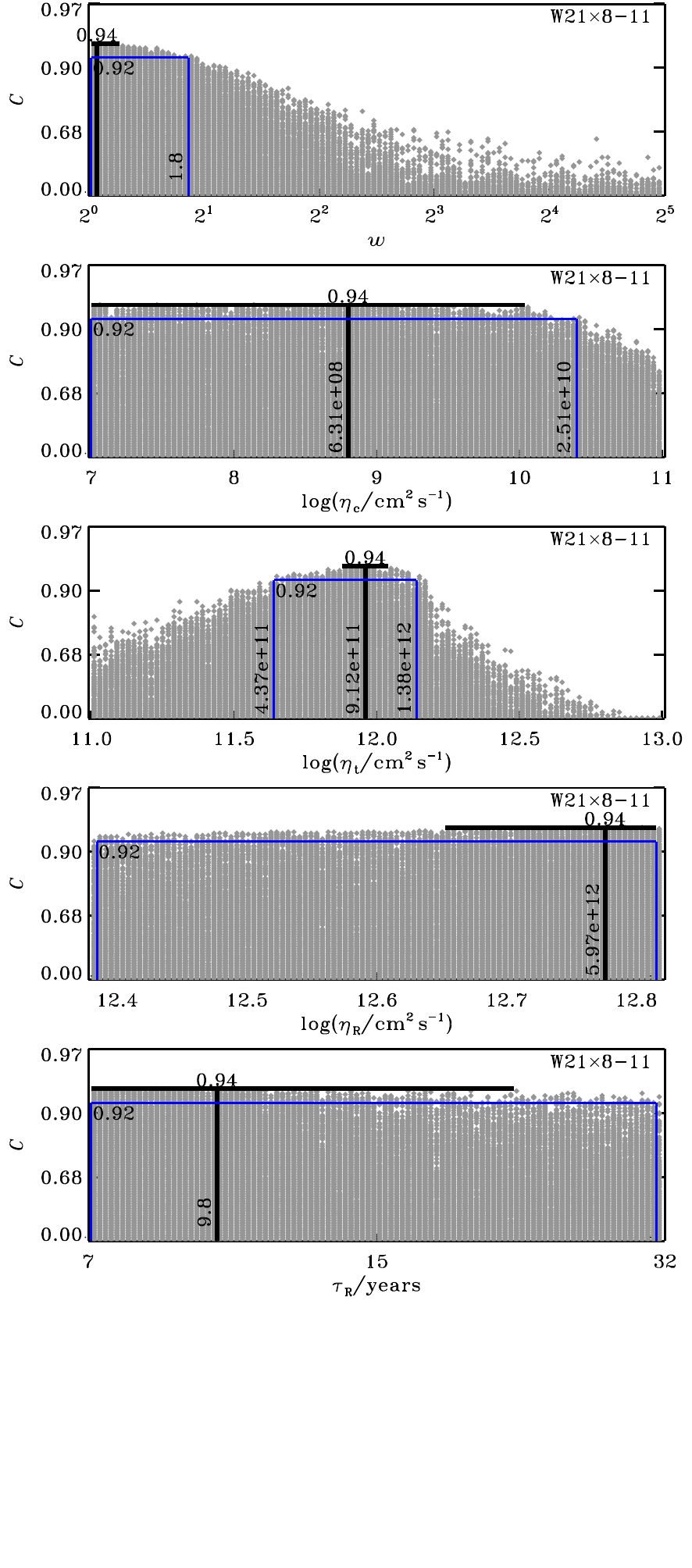}
   \caption{
   Same as Figure~\ref{f_fitland0}, but for the eleven model parameters 
   ($\delta_c$, $R_b$, $u_0$, $p$, $q$, $v$, $w$, $\eta_c$, $\eta_t$, 
   $\eta_R$, and $\tau_R$), from four independent W21$\times$8-11 
   optimisations, while the seven ``emergence'' parameters are held 
   fixed to their optimal value listed in Table~\ref{t_param}.
   }
   \label{f_fitland}
\end{figure*}

The first sequence of optimizations are run with all $18$ unconstrained 
parameters allowed to vary freely in the intervals listed in 
Table~\ref{t_param}, hence called W21$\times$8-18.
We first analyse the model's behavior relative to the parameters 
involved in the very definition of the emergence function 
$F_B(\theta,t)$ (Equation~(\ref{eq_fitfunc})).
Figure~\ref{f_fitland0} illustrates the value of the goodness-of-fit 
$C$ as a function of emergence parameters $r^\ast$, $b$, $a$, $c$, 
$B^\ast$, $\ell^\ast$, and $\mu_\ell$ for a set of $192000$ solutions 
obtained from four independent optimizations (different seed 
populations), $500$ generations each, $96$ trial solutions per 
generation. 
In all four optimizations, the fitness reaches the same optimal value 
$C_\text{max}=0.94$. 
Such optimal solution, which parameters are listed in bold font in the 
rightmost column of Table~\ref{t_param}, is presented in 
Figures~\ref{f_W21x8}(e) and \ref{f_W21x8}(f). 
The fit between the emergence function (Figure~\ref{f_W21x8}(f)) and the 
smoothed butterfly diagram of cycle 21 emergences 
(Figure~\ref{f_W21x8}(h)) is good, with expected butterfly shapes and 
cycle overlaps. 

However, it is clear from Figure~\ref{f_fitland0} that considering only 
a single optimal solution is insufficient, optima being surrounded by 
a wide variety of sub-optimal but likely acceptable solutions, besides 
the clearly unacceptable ones.
Also, all seven parameters presented are not equally constrained by 
the fitting procedure. By looking at all solutions standing above the 
$C\geq0.935$ level (thick black line), we get a first estimate of the 
relative restriction applied on each parameter. 
For instance, parameters $r^\ast$, $a$, and $B^\ast$ are fairly 
well constrained to a limited interval within the original boundaries, 
while parameters $b$, $c$, $\ell^\ast$, and $\mu_\ell$ show wider 
regions of acceptable fit.

In order to build meaningful error estimates for each parameter, we 
must assess the physical limit of validity of the optimization 
criterion. 
Clearly, there must exist a value of $C$ above which solutions are 
physically acceptable, even if not strictly optimal. 
An example of such a solution, with $C=0.92$, is presented in 
Figures~\ref{f_W21x8}(c) and \ref{f_W21x8}(d). 
The butterfly shape in this solution is still clearly visible, though a 
second tail is starting to build towards the high latitudes.
These differences are significant enough to declare such a solution 
inferior to the optimal one, but still at the limit of acceptability 
in terms of observed global features.
The horizontal blue lines in Figure~\ref{f_fitland0} delimit the 
solutions that are characterized by a criterion $C\geq0.92$.

Before proceeding further into the parameters analysis, we now opt to 
get rid of the variability associated with the definition of the 
empirical emergence function (Equation~(\ref{eq_fitfunc})), and pick up 
definitive values, within the interval of acceptability, for the 
parameters involved. 
The inferred depth for the generation of flux instabilities is thus set 
near its optimal value $r^\ast/R=0.68$, by averaging the 
magnetic field values between $r/R=0.68$ and $0.70$. 
For simplicity, the relative contribution to $B_\text{mix}$ of the 
poloidal field is set to zero ($a=0$), while we round the optimal 
exponent of the toroidal contribution to $b=1.5$. 
The lower threshold, above which this diffuse toroidal field is assumed 
to be able to generate instabilities, is set to its highest acceptable 
value, that is $B^\ast=10^2$. 
The units of $B^\ast$ are in fact $\gauss^{1.5}$ in the case $b=1.5$ to 
ensure coherence in Equation~(\ref{eq_fitfunc}). 
This corresponds to a lower threshold of $\simeq\unit{200}{\gauss}$ in 
$B_\phi$, as illustrated in Figure~\ref{f_stab}. 
The emergence function $F_B$ remains proportional to $B_\text{mix}$, 
with $c=1.0$, rather than saturating above $B^\ast$. 
The highest latitude of emergence is fixed to 
$\ell^\ast=\unit{70}{\degree}$ ($\sin\ell^\ast=0.94$), in accordance 
with stability diagrams by \citet{FerrizMas1994}, and the equatorial 
intercept $\mu_\ell$ is set to $0.5$, such that the latitudinal filter 
halves smoothly from $\ell^\ast=\unit{70}{\degree}$ down to the equator. 
The final emergence function (i.e. emergence probability) can now be 
mapped as a function of latitude and toroidal field amplitude, as shown 
in Figure~\ref{f_stab}, to form a synthetic ``stability'' diagram, 
which is the model's equivalent to the stability diagrams presented in 
\citet[Figures 1 and 2]{FerrizMas1994}.

\begin{figure}
   \plotone{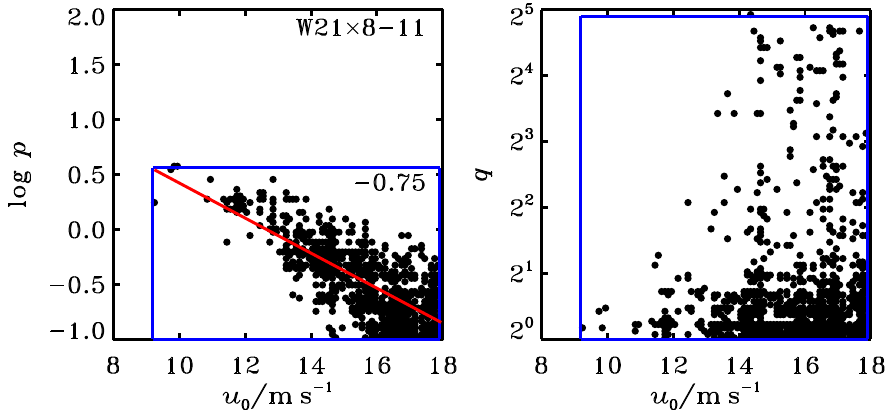}
   \caption{Correlations between the best-fit parameter values for 
   surface meridional flow speed $u_0$ and (left) depth variation 
   parameter $p$ and (right) latitudinal profile parameter $q$
   (see Equations~(\ref{eq_mercirca}) and (\ref{eq_mercircb})). 
   The blue squares correspond to the $C\geq0.92$ regions on 
   Figure~\ref{f_fitland}, third, fourth, and fifth panels at left. 
   The linear best-fit (red line) and Pearson's correlation coefficient 
   are also shown on the left panel. 
   In particular, despite the ranges of values for $u_0$, $p$, and $q$, 
   all these solutions have a peak equatorward flow speed of 
   $\unit{6.6}{\mps}\pm8\%$ near $r/R=0.66$.}
   \label{f_correl}
\end{figure}
With the emergence function now fixed, we carry out a new series of 
four optimizations, hereafter called W21$\times$8-11, with only the 
$11$ physical model parameters ($\delta_c$, $R_b$, $u_0$, $p$, $q$, 
$v$, $w$, $\eta_c$, $\eta_t$, $\eta_R$, and $\tau_R$) left to vary 
freely in their prescribed intervals.
The corresponding $192000$ solutions are presented in 
Figure~\ref{f_fitland} as a function of each parameter values. 
Again, the optimal fitness reaches $C_\text{max}=0.94$, and all 
solutions characterized by a $C\geq0.92$ are considered acceptable. 
The corresponding interval for each parameter is used to define final 
error bars about the optimal values, as listed in the rightmost column 
of Table~\ref{t_param}. 
As mentioned earlier, various combinations of parameters within these 
accepted intervals lead to acceptable solutions, but not all do, due
to various correlations between some pairs of best-fit parameters 
(see also discussion in Paper~I, \S~3.5). 
Figure~\ref{f_correl} depicts two of the strongest such correlations 
uncovered in our W21$\times$8-11 set of solutions. 
The left panel shows a net linear (anti)correlation between the surface 
meridional flow speed $u_0$ and one of the parameters ($p$) setting the 
depth dependence of the meridional flow in the interior 
(viz. Equation~\ref{eq_mercirca}). 
This (anti)correlation has an unambiguous physical explanation: 
it leads to all solutions near the red line having an equatorward 
meridional flow speed equal to $\unit{6.6}{\mps}\pm8\%$ at $r/R=0.66$, 
that is below the base of the convective envelope, beneath the 
layer where the emergence function is calculated. 
It is the speed of this return flow that sets the cycle period,
and thus is strongly constrained by the sunspot butterfly diagram used 
to establish our goodness-of-fit measure. 
The right panel of Figure~\ref{f_correl} shows another correlation 
between a pair of parameters, in the form of a somehow triangular 
constraint on parameter $q$, which controls the polar end of the 
latitudinal dependence of the meridional flow, as a function of maximum 
flow speed $u_0$ (viz. Equation~\ref{eq_mercircb}).
This correlation sets a lower limit on the surface flow speed at 
mid--high latitude, of the order of $\gtrsim\unit{5}{\mps}$.


\section{A Solar-like Dynamo Solution} \label{s_bldyn}

Now that the physical model and masking parameters have been properly 
calibrated to ensure that function $F_B(\theta,t)$ reproduces the 
observed solar butterfly diagram of surface emergences, we may use it 
as the statistical emergence function it was meant to be, i.e. 
providing the missing surface source term $S_\text{BMR}(\theta,\phi,t)$ 
with new emergences generated from deep seated toroidal flux 
(Equations~(\ref{eq_sbmr}) and (\ref{eq_bbmr})) and thus closing 
the loop for a self-consistent and autonomous $2\times2$D dynamo.

In all following cases, we use as initial condition the simulation 
state at the end of the previously calibrated W21$\times$8 sequences.
This ensures that the new simulations start up from a state 
representative of a solar activity minimum.

\subsection{Quasi-Linear Regime} \label{s_bldynlin}

\begin{figure}
   \plotone{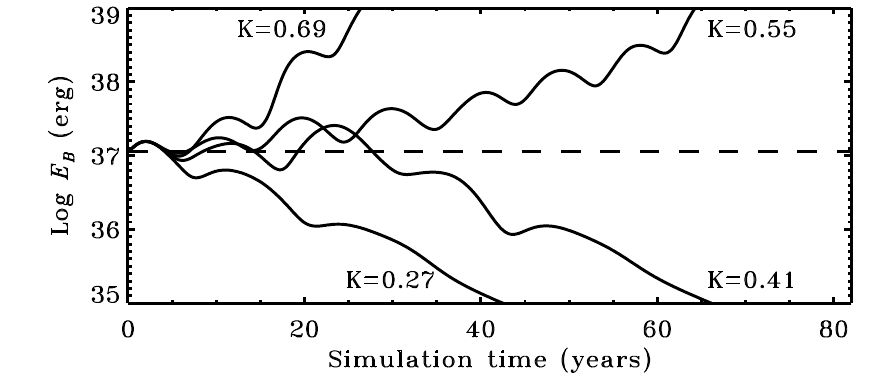}
   \plotone{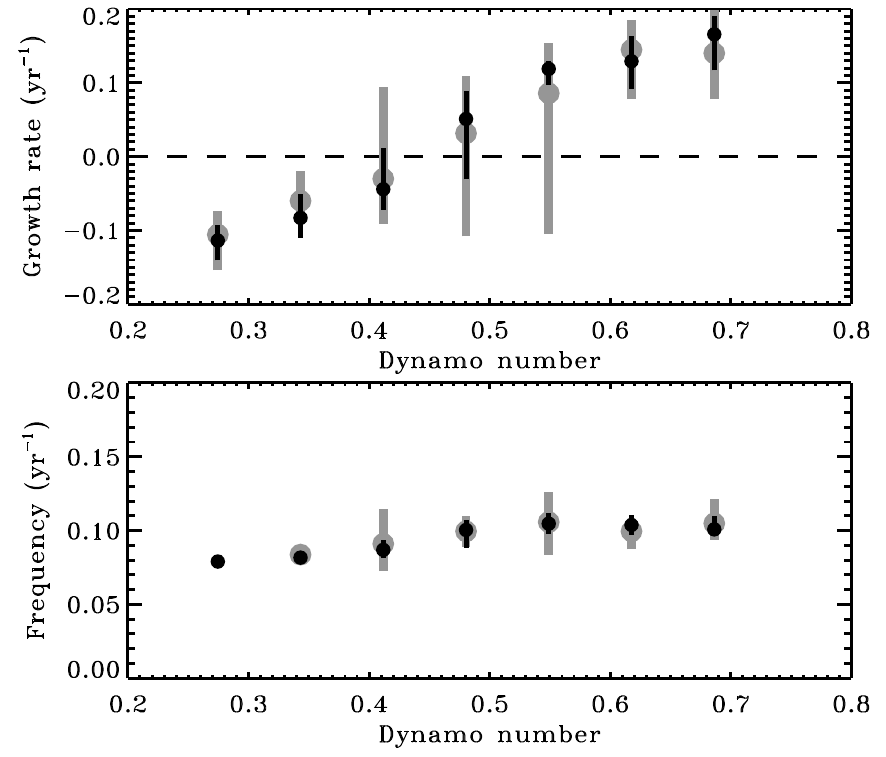}
   \caption{Top panel: evolution of the total magnetic energy content 
   inside the simulated Sun, for $\simeq8$-cycles sample 
   realizations of a $2\times2$D dynamo run in the quasi-linear regime 
   at four different dynamo numbers $K$ (horizontal dashed line 
   indicates the initial energy level). 
   Middle panel: long term growth rate of the magnetic energy as a 
   function of dynamo number $K$, for ten independent realizations of 
   (thick gray) the full statistical emergence procedure 
   (cf. \S~\ref{s_emerg}, stochasticity sources (i) to (vi)) 
   per value of $K$ and of (thin black) a reduced stochastic emergence 
   procedure (retaining sources (i) to (iv) only, and fixing bipole 
   separations (v) and tilts (vi) at their observed mean values). 
   Bottom panel: similar as the preceding panel, but for the oscillation 
   frequency of the detrended magnetic energy.}
   \label{f_growthfreq}
\end{figure}
\begin{figure*}
   \plotone{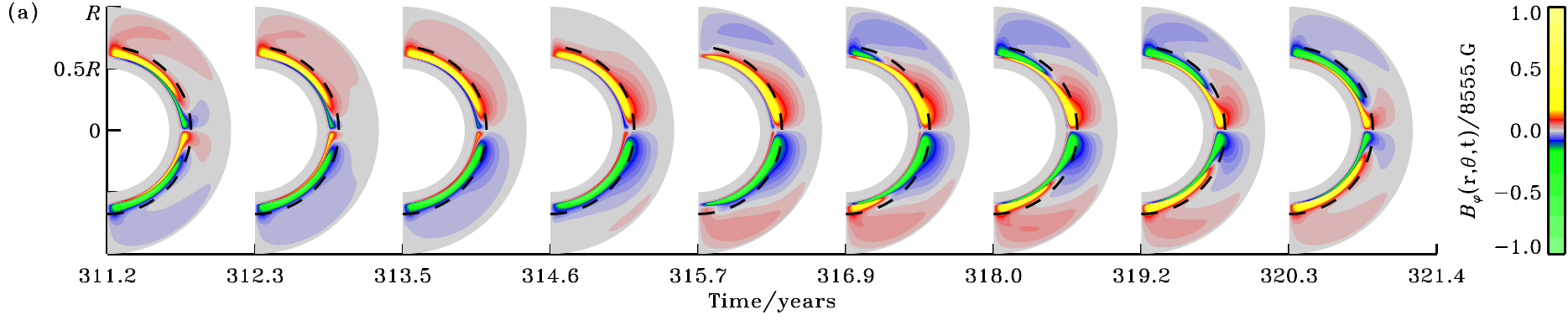}
   \plotone{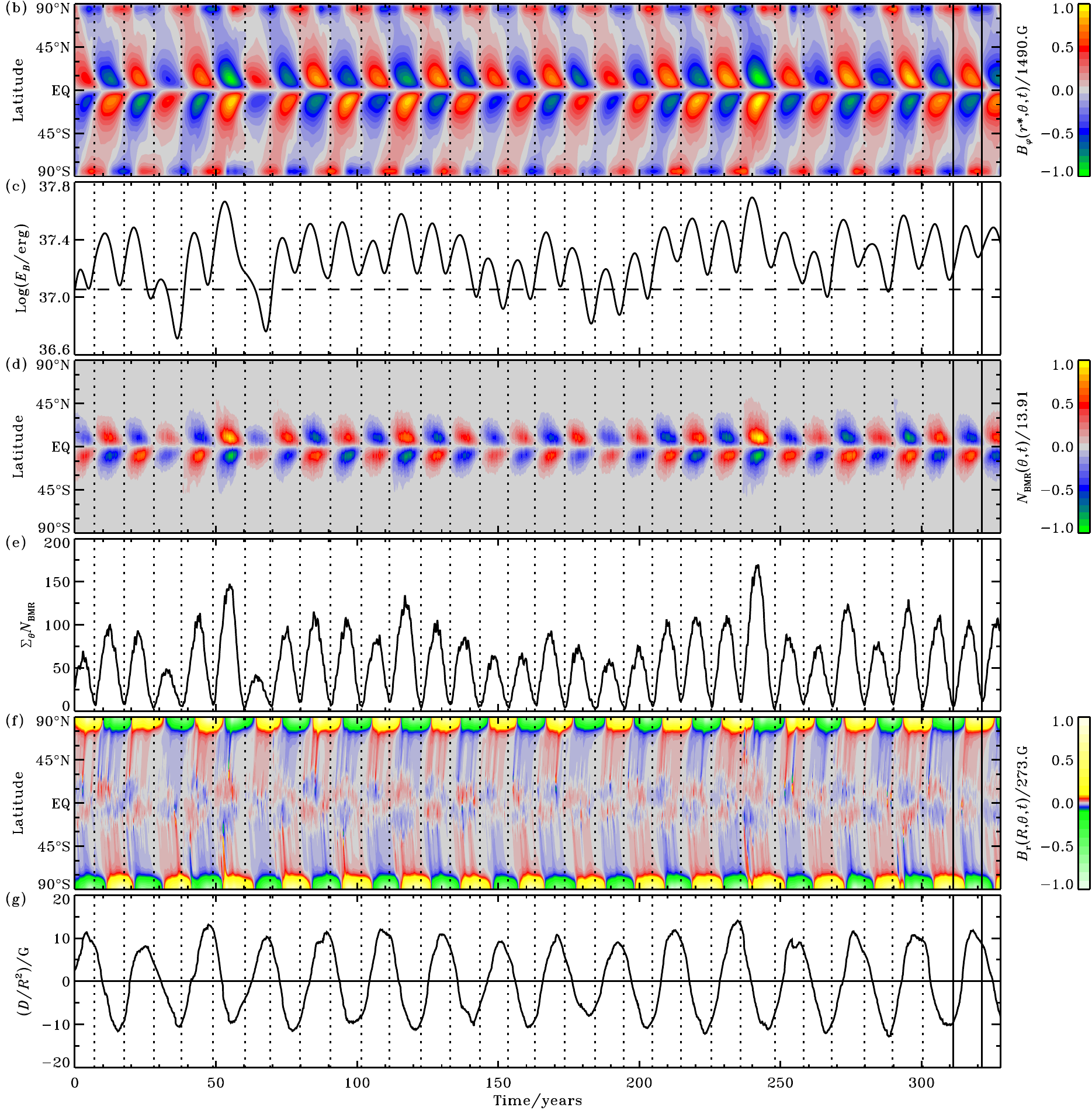}
   \caption{
   A representative solar-like tilt-quenched $2\times2$D dynamo solution 
   obtained using the optimal parameter values listed in the rightmost 
   column of Table~\ref{t_param}. 
   (a) Latitude--radius snapshots of the toroidal magnetic field between 
   $r/R=0.5$ and $1.0$, at nine different phases of the dynamo cycle 
   delimited by two vertical continuous lines in the following plots 
   (color table saturates above $\unit{1.5}{\kilo\gauss}$; 
   dashed lines indicate the depth of the tachocline ($r/R=0.7$)). 
   (b) Time--latitude contour plot of the toroidal magnetic field 
   averaged in the depth range $0.68\leq r^\ast/R\leq0.70$; 
   (c) corresponding temporal evolution of the total magnetic energy 
   content inside the simulated Sun ($0.5\geq r/R\geq1.0$; horizontal 
   dashed line indicates the initial energy level). 
   (d) Time--latitude density plot (butterfly diagram) of the number of 
   \acp{BMR} emerged at the surface, as dictated by the emergence 
   function $F_B$, in turn based on the preceding toroidal field 
   amplitude; 
   (e) corresponding monthly number of newly emerged \acp{BMR} 
   (pseudo-SSN), as a function of time. 
   (f) Time--latitude contour plot of the surface radial magnetic field 
   (color scale saturated above $\unit{27}{\gauss}$); 
   (g) corresponding temporal evolution of the surface axial dipole 
   moment. 
   Vertical dotted lines indicate the times of activity minima as 
   defined by the minimum values of the pseudo-SSN.
   }
   \label{f_refdyn}
\end{figure*}
The linearity in $\vec{B}$ of the \ac{FTD} 
equations~(\ref{eq_mhdaxia}) and (\ref{eq_mhdaxib}) and 
\ac{SFT} equation~(\ref{eq_surftrans}) 
is expected to lead to either growing or decaying dynamo solutions. 
In the well-studied mean-field framework, this behavior is controlled 
by the ``dynamo number''. 
Here it is the proportionality constant $K$ between $F_B(\theta,t)$ 
and the absolute number of emerging \acp{BMR} per time step that 
plays the equivalent role. 
Since $a=0$ and $b=1.5$ in the definition of $B_\text{mix}$, and $c=1$ 
in Equation~(\ref{eq_fitfunc}), the number of emerging \acp{BMR} is 
proportional to ${B_\phi}^{1.5}$ as long as the latter exceeds the 
lower threshold $B^\ast$. 
However, the emergence process itself is inherently stochastic, 
so the dynamo growth rate can only be defined in a statistical sense, 
hence the ``quasi-linear'' labeling.

The top panel of Figure~\ref{f_growthfreq} depicts the temporal 
evolution of the total magnetic energy content inside the simulated Sun, 
for $\simeq8$-cycles sample realizations of a $2\times2$D dynamo 
run in the quasi-linear regime at four different dynamo numbers $K$.
From these few samples, the transition between decaying (small $K$) and 
exponentially growing (large $K$) solutions seems sharp, 
but a more complete analysis reveals otherwise. 
The middle panel of Figure~\ref{f_growthfreq} shows how the growth rate 
of the magnetic energy can show a wide spread at a given value of $K$.
Error bars on the plot illustrate the intervals of growth rates obtained 
at each given $K$, through ten different realizations of the statistical 
emergence procedure described earlier (cf. \S~\ref{s_emerg}). 
We also performed a similar set of simulations in a reduced stochastic 
regime (shown in black on the plot). 
This reveals the strong global impact of stochasticity in the emergence 
process, particularly by the distributions in separations and 
tilts of emerging \acp{BMR}. 
The consequence is that a precise value for the critical dynamo number
cannot be defined, with different realizations of the dynamo with 
$K\in[0.4,0.6]$ resulting in dynamo solutions that can either grow or 
decay. 
The fact that this transition region lies significantly below the value 
$K\simeq1$ required to reproduce the observed butterfly diagram for 
cycle 21 in the preceding section suggests that the dynamo should run 
in the supercritical regime, with some non-linear feedback regulating 
the mean cycle amplitude. 
This aspect will be discussed in the following subsection. 

As another indicator of the model behavior, average cycle frequencies 
(periods) of the corresponding solutions, are also presented in the 
bottom panel of Figure~\ref{f_growthfreq}, again with error bars showing 
the intervals of frequencies obtained for a given $K$. 
Considering the difficulty of measuring cycle periods in quickly 
decaying oscillatory solutions (low $K$), no strong trend appears from 
this plot. 
This suggests how robust is the model at producing oscillations on a 
$\unit{9-12}{\years}$ timescale, in spite of the strong variability 
associated with stochastic processes.


\subsection{Tilt-Quenching and Reference Dynamo Solutions} \label{s_bldynqch}

To overcome the problem of (quasi-)linearity, but without dealing 
explicitly with dynamical feedback, some ad hoc quenching may be added 
to the dynamo source terms. 
Motivated by the 
modeling of the buoyant rise of thin magnetic flux tubes by 
\citet{DSilva1993} and \citet{Caligari1995} 
(for a review see \citealt{Fan2009}, \S~5.1.2, and references therein), 
we introduce a quenching of the \ac{BMR} mean tilt as a function of
the amplitude of the contributing underlying toroidal field 
$B_\phi(r^\ast,\theta,t)$, in order to mimic the resistance of magnetic 
tension in strongly magnetized flux tubes against the twisting imparted 
by the Coriolis force. 
Observationally the situation is less clear-cut
(see \S~6 in \citet{Pevtsovetal2014} for a recent review).
\citet{DasiEspuig2010} and \citet{McClintockNorton2013} do find an 
influence of cycle amplitude on mean tilt angles, varying from cycle 
to cycle and from one solar hemisphere to another, but 
\citet{Stenflo2012} do not find a statistically significant 
relationship between tilt angles and flux of individual \acp{BMR}.

The quenched tilt is written 
\begin{equation}
   \alpha_q = \frac{\alpha}{1+(B_\phi/B_q)^2} \ ,
\end{equation}
with $B_q$ some ajustable critical magnetic field amplitude.
In the context of the present dynamo model, we find a tilt-quenching 
with $B_q\simeq\unit{500}{\gauss}$, at dynamo number $K=0.75$, to be 
adequate to generate stable dynamo solutions, comparable to solar 
amplitudes for the butterfly density plot and the monthly number of 
newly-emerged \acp{BMR}. 
The latter we refer to as a ``pseudo-SSN'', since no consideration is 
given here to distinguishing groups vs individual emergences, or 
assigning them different weights, as is the case in the definition of 
the international SunSpot Number (SSN). 
For instance, observed cycle 21 peaks at a SSN of $\simeq175$ while the 
maximum monthly number of newly-emerged \acp{BMR} in 
\citet{Wang1989-0}'s database is $\simeq50$. 

Figure~\ref{f_refdyn}(b)--(g) illustrate the evolution of the deep 
toroidal field, total magnetic energy, \acp{BMR} density, pseudo-SSN, 
surface radial field, and axial dipole moment for a sample 
dynamo solution run over more than $300$ years and roughly $32$ 
synthetic solar cycles. 
The temporal series exhibit solar-like behaviors in many aspects, 
in particular cycle periods varying between $8.5$ and $12$~years, 
cycle amplitude variations of a factor three to four in the 
pseudo-SSN, and long term variability such as some progressive increase 
of cycle amplitude after the occurence of a weak cycle or the triggering 
of small cycles after very strong ones. 
Some significant hemispheric asymmetries are also noticeable on the 
various plots, but polarity reversals remain sharply synchronized, 
indicating strong cross-hemispheric coupling. 
The oscillating surface axial dipole moment peaks at or near pseudo-SSN 
minimum, in agreement with observations. 
The phase relationship between the surface dipole and deep-seated 
toroidal field is also solar-like, with the dipole peaking 
at or shortly prior to pseudo-SSN minimum. 

The overall amplitude of this dynamo solution is however 
slightly higher than that of the average solar cycle. 
The axial dipole moment (panel (g)) oscillates with an amplitude of 
$\simeq\unit{10}{\gauss}\cdot R^2$ as compared to 
$\simeq\unit{4}{\gauss}\cdot R^2$ for the Sun.
The pseudo-SSN (panel (e)) peaks between $\simeq50$ and 
$\simeq150$, which is slightly higher than average solar cycle amplitude 
($\simeq50$ for cycle 21). 
This corresponds to $50-150$ emergences per month near cycle maxima 
down to $0-10$ per month at cycle minima. 
The total number of \acp{BMR} to emerge during a cycle varies from 
$\simeq2000$ for the smallest cycles to $\simeq8000$ for the strongest 
ones, which is comparable, but again slightly higher in average, than 
the original $\simeq3000$ \acp{BMR} extracted from \citet{Wang1989-0}'s 
database for cycle 21. 
Due to the use of a constant log-normal distribution for \ac{BMR} 
magnetic fluxes, total magnetic flux emerged during a cycle scales 
linearly with the number of \acp{BMR}. 
Finally, most presumably due 
to the use of a suboptimal profile for the surface meridional 
circulation leading to extra flux accumulation near the poles at 
activity minima, the peak amplitude of the radial surface field 
(panel (f)) 
builds up at an order of magnitude stronger than observed. 
At any rate, a dynamo number $K\lesssim1$ is more than sufficient to 
maintain stable dynamo solutions, with only the \ac{BL} mechanism 
operating and without having to artificially enhance the emerging flux 
(see also \citealt{CameronSchussler2015}). 
The value of $K$ used here for the reference dynamo should even be 
brought down a little to better fit solar cycle observations.

Also shown in Figure~\ref{f_refdyn}(a) is a series of radius--latitude 
cuts of the toroidal field component, at nine different phases of a 
synthetic sunspot cycle. 
The toroidal field reverses amplitude after $\simeq\unit{9}{\years}$, 
which is slightly shorter than the average observed sunspot cycle. 
The peak toroidal field amplitude near $r/R=0.7$ is reached at 
mid-cycle, near maximum sunspot activity. 
Below the tachocline, the magnetic field from three to four successive 
cycles piles up to thinner and thinner layers as it reaches the depth 
$r/R=0.6$.
This is precisely what is to be expected from the average diffusivity 
$\eta\simeq\unit{5\times10^{10}}{\cmmps}$ used at $0.6\geq r/R\geq0.7$, 
which corresponds to a diffusive time-scale of $\simeq31$~years. 
Below $r/R=0.6$, the magnetic diffusivity of $\unit{10^9}{\cmmps}$, 
leads to a diffusive time-scale $\gtrsim1000$~years. 
Therefore, while the meridional circulation acts on a time-scale 
commensurate with the sunspot cycle period, the deep diffusive processes 
act on much longer timescales.
The remnants from old cycles appear to be able to feed back into the 
dynamo system and induce some long term memory in cycle amplitude.

\begin{figure}
   \plotone{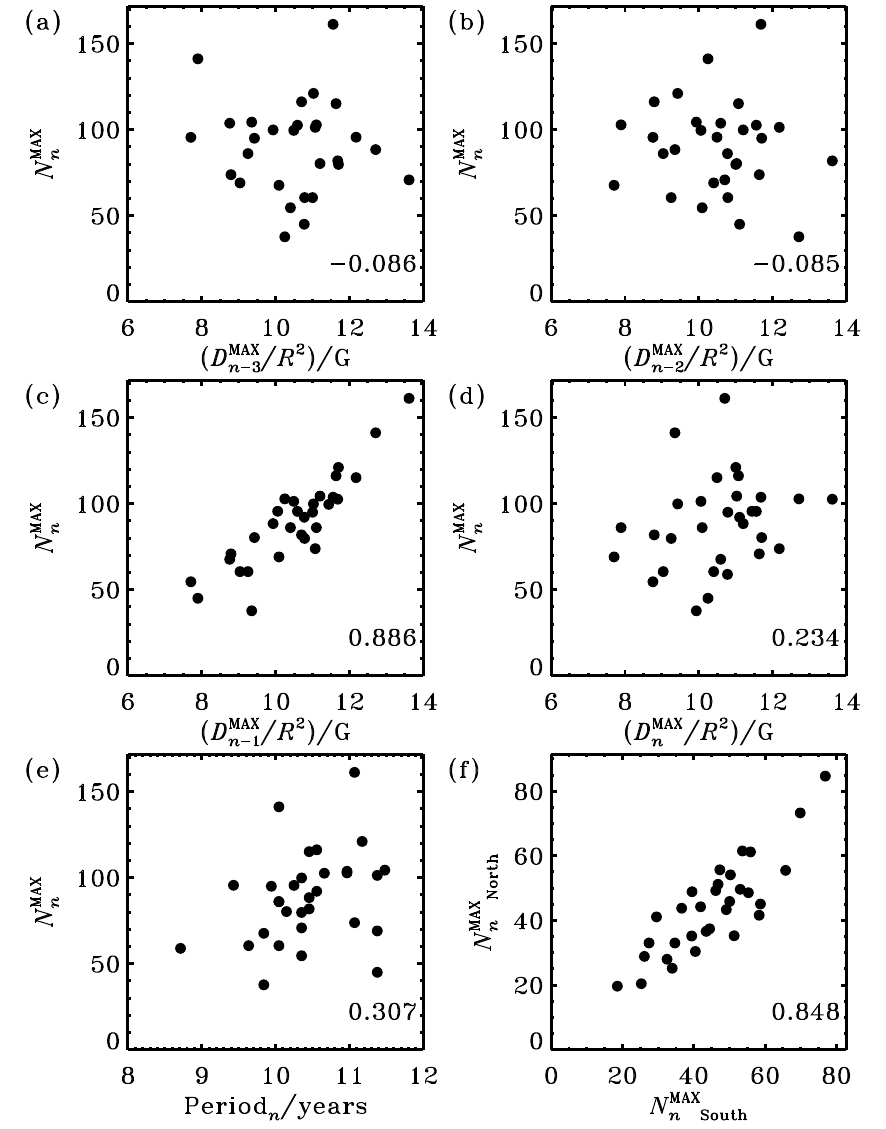}
   \caption{
   Amplitude (maximum pseudo-SSN) of cycle $n$ as a function 
   of maximum axial dipole moment at the end of (a) cycle $n-3$, 
   (b) cycle $n-2$, (c) cycle $n-1$, and (d) cycle $n$, for the 
   sample dynamo solution presented in Figure~\ref{f_refdyn}. 
   (e) Cycle amplitude as compared to the period of the same cycle. 
   (f) Cycle amplitude calculated independently in each hemisphere and 
   plotted against one another. 
   In each panel is also given the corresponding Pearson's linear 
   coefficient.}
   \label{f_refdyncorrel}
\end{figure}
Figure~\ref{f_refdyncorrel} shows some long term interrelations between 
cycle properties, extracted from the preceding dynamo solution. 
Panel (c) in the figure shows the strong linear 
correlation ($0.89$) obtained between amplitude (maximum pseudo-SSN) 
of a cycle ($n$) and maximum axial dipole moment at the end of 
the preceding cycle ($n-1$). 
This behavior is to be expected from the quasi-linear transport and 
shearing of the poloidal magnetic field accumulated at cycle minimum 
into a deep toroidal component peaking at cycle maximum and generating 
a proportional number of surface emergences. 
As shown in panel (d) of the figure, the reverse 
correlation is not true, however, as the stochastic properties of 
emerged \acp{BMR} during a given cycle $n$ destroy the otherwise
expected correlation between pseudo-SSN and axial dipole amplitude at 
the end of the same cycle ($n$). 
Also, even if long term magnetic memory does exist in the interior, the 
poor correlations obtained between amplitude of cycle $n$ and axial 
dipole moment at the end of cycles $n-2$ (panel (b)) and cycles $n-3$ 
(panel (a)) indicate that it is erased by the stochasticity of flux
emergence. 
Despite these stochastic sources of fluctuations, hemispheric cycle 
amplitudes remain strongly correlated, as shown in 
Figure~\ref{f_refdyncorrel}(f).
All the preceding results are in good agreement with observed solar 
cycle characteristics (see, e.g., \citealt[Figure~5]{Munoz2013}).

As also shown in panel (e) of Figure~\ref{f_refdyncorrel}, 
cycle amplitude and period are essentially uncorrelated. 
This differs from the behavior observed in the Sun, where a significant 
anticorrelation is inferred between these two cycle measures. 
Some additional dynamical feedback would likely be required to reproduce 
such behavior.

\subsection{Long Term Variability}

\begin{figure}
   \plotone{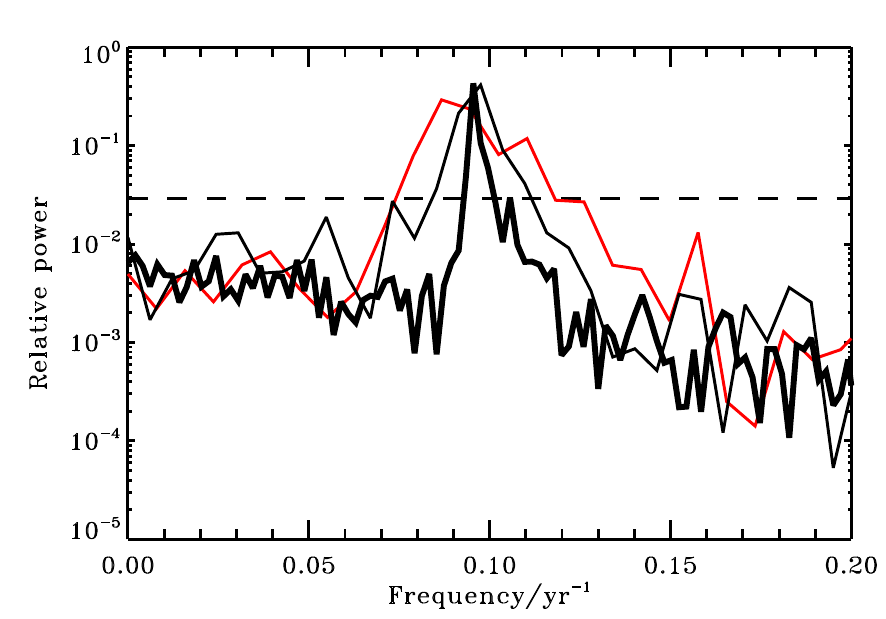}
   \caption{
   Temporal power spectra of the solar SSN (thin red), 
   of the pseudo-SSN of a sample 32-cycles tilt-quenched 
   $2\times2$D dynamo simulation (thin black), 
   and of the pseudo-SSN averaged over three independent realizations 
   of a 96-cycles simulation (thick black). 
   In all three cases, the Fourier transform was performed on a 
   signed version of the temporal series, with amplitudes alternatively 
   reversed from one cycle to the next to impose an oscillation about 
   zero, and the frequencies subsequently multiplied by two to 
   retrieve the $\simeq\unit{10}{\years}$ sunspot cycle characteristic 
   period.
   The horizontal dashed line marks $\unit{10}{\%}$ of the peak 
   spectral power.
   }
   \label{f_dynfft}
\end{figure}
\begin{figure*}
   \plotone{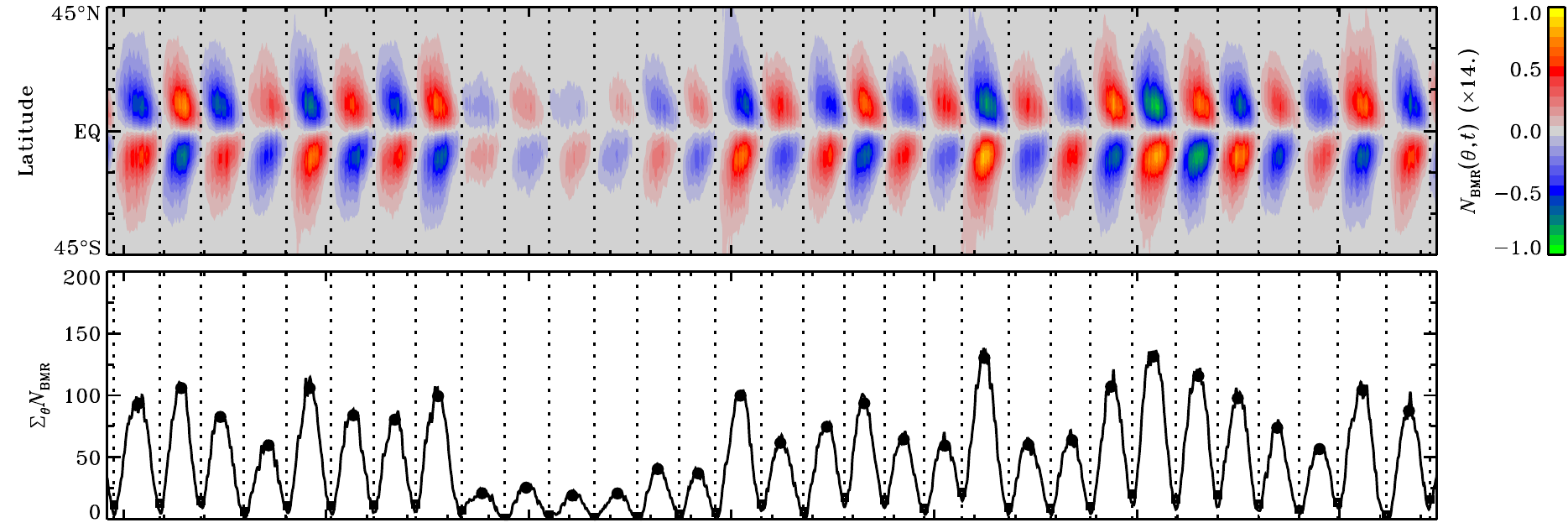}
   \plotone{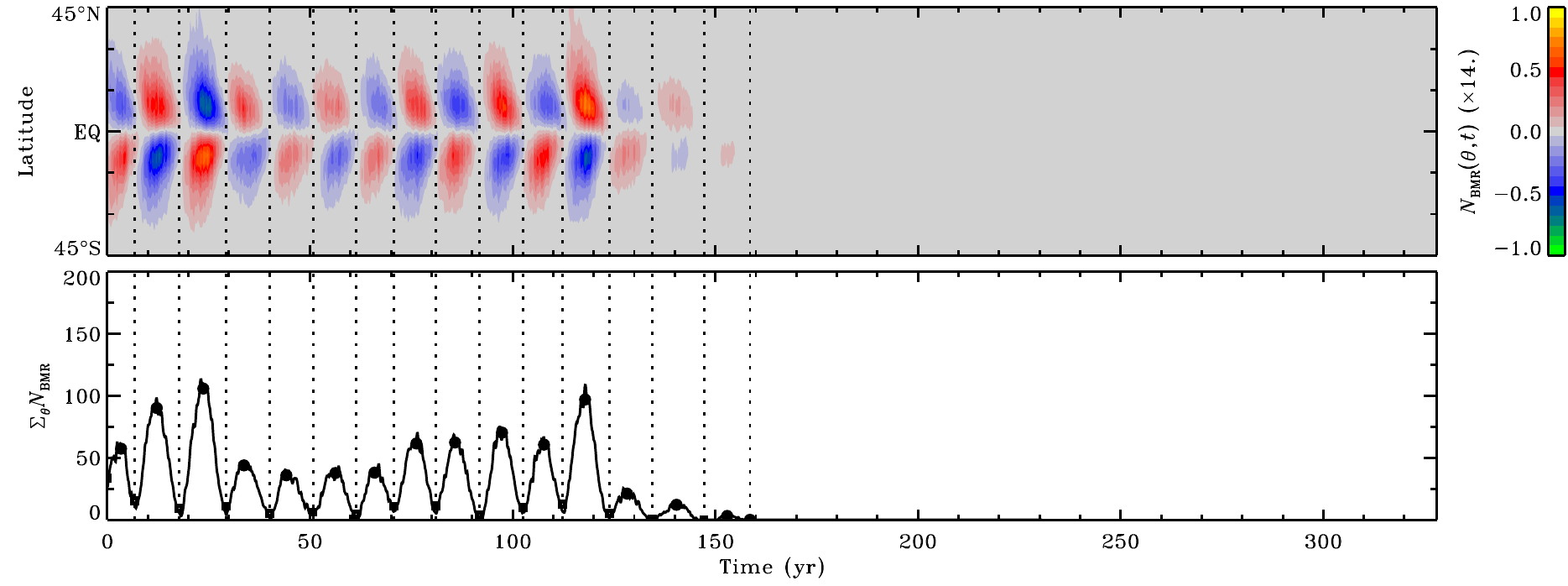}
   \caption{Time--latitude density plot (butterfly diagram) and 
   corresponding monthly number (pseudo-SSN) of newly emerged \acp{BMR}, 
   for two distinct realizations of a 32-cycles 
   tilt-quenched $2\times2$D dynamo simulation using the same 
   optimal parameter values used to produce the reference solution 
   of Figure~\ref{f_refdyn} (viz. panels (d) and (e)).}
   \label{f_dyn3sol}
\end{figure*}
Figure~\ref{f_dynfft} shows the Fourier transforms of the pseudo-SSN 
time series, for a sample 32-cycles tilt-quenched $2\times2$D dynamo 
simulation similar to the reference solution of Figure~\ref{f_refdyn}, 
along with the average spectra constructed from three statistically 
independent realizations of a 96-cycles simulation. 
The relatively poor sampling of the 32-cycles simulation shows spectral 
features similar to those of the 23-cycles solar SSN spectrum (also 
shown in the figure), in that it presents a broad peak between periods 
of $9$ to $\unit{12}{\years}$ ($8$ to $\unit{14}{\years}$ for the SSN) 
as well as low amplitude ($\unit{5-10}{\%}$ of peak power) structures 
at other frequencies. 
However, these secondary features occur at different frequencies for 
the SSN and for different realizations of the pseudo-SSN, and so do not 
represent physically robust signals. 
Indeed, the averaging of three 96-cycles spectra (equivalent to 
$\simeq300$ cycles in total) reveals no hint of low-frequency signature 
above $\unit{2}{\%}$ of peak power, of the type one would associate with 
the so-called Gleissberg or Suess cycles detected in temporally-extended 
records of solar activity. 
The cycle period is also much more robust, at $\unit{9.5-11}{\years}$. 
These results indicate that despite the strong variability in cycle 
amplitude characterizing the simulations, the period is very stable, 
even more so than in the real Sun. 

Figure~\ref{f_dyn3sol} shows two sets of synthetic butterfly diagrams 
and associated pseudo-SSN time series, obtained for the same parameter 
values as the solution of Figure~\ref{f_refdyn} but using distinct 
stochastic realizations for the fluctuating properties of the synthetic 
\acp{BMR}. 
The top solution generally resembles panels (d) and (e) of 
Figure~\ref{f_refdyn} in its overall amplitude fluctuation pattern, but 
now also shows an episode of strongly reduced cycle amplitude, 
persisting here for four cycles ($84\leq t\leq\unit{132}{\years}$) and 
reminiscent of the 1796--1825 Dalton minimum of the sunspot record. 
Entry into this low amplitude episode is sudden, the preceding few 
cycles being of average amplitude or higher. 
Recovery is however more gradual, with a few cycles required 
for the cycle to build back up to its pre-event average amplitude.

The solution plotted on the two bottom panels of Figure~\ref{f_dyn3sol} 
shows yet another interesting behavior: 
a complete halt of the cyclic dynamo, here at 
$t\simeq\unit{150}{\years}$, following a sequence of unfavorably 
positioned and/or tilted large \acp{BMR}, leading to a much reduced 
dipole moment building up in the descending phase of the cycle 
peaking at $t\simeq\unit{118}{\years}$. 
Because of the lower cutoff $B^\ast$ built into our emergence function 
(viz. Equation~(\ref{eq_fitfunc}) herein), once the toroidal magnetic 
field falls below this threshold, \acp{BMR} are no longer produced, so 
that the existing dipole then undergoes simple resistive decay, 
followed by resistive decay of the toroidal component, 
as per Cowling's theorem. 
A distinct inductive mechanism able to operate at low mean-field 
strengths, such as the alpha-effect of classic mean-field 
electrodynamics, would be needed here to restart the dynamo cycle
(see, e.g., \citealt{Passosetal2014}).
Ongoing numerical experiments along these lines suggest that this would 
be a feasible path towards the generation of solar-like Grand Minima of 
activity.

In a set of $30$ realizations similar to the one displayed in 
Figure~\ref{f_refdyn} and the two in Figure~\ref{f_dyn3sol}, seven 
shut off before reaching the $32$th cycle, and 15 before reaching 
the $96$th cycle. 
The probability of a dynamo to remain active after a certain number of 
cycles thus decreases with time in a manner that appears consistent 
with a stationary memoryless random process, as would be expected from 
the stochastic nature of the properties of emerging \acp{BMR} built 
into the model. 
A detailed, quantitative investigation of these matters, currently 
underways, will be the focus of a subsequent paper in this series.

\section{Discussion}

The dynamo solutions presented above result from the use of a model 
calibrated to cycle 21 emergence data through an optimization process 
operating on a specific goodness-of-fit measure and in a bounded search 
space. 
These bounds were set (loosely) on observational and/or physical 
grounds, but obviously pose a restriction on the range of solutions 
accessible to the optimization. 
Could we do better than the optimal solution listed in 
Table~\ref{t_param}~? 
We have carried out a number of alternate optimization runs in order to 
answer this question, as described in what follows.

An 18-parameter optimization similar to that described in 
\S~\ref{s_optsol} but using much broader ranges of parameter does manage 
to return a best-fit solution with $C=0.97$, significantly better than 
the original 18-parameter best-fit solution, which has $C=0.94$. 
This nominally superior fit, however, is achieved through a low-latitude 
cutoff for the emergence function, down to 
$\ell^\ast=\unit{30}{\degree}$, which is clearly incompatible with 
stability diagrams for thin toroidal flux ropes.

\begin{figure}
   \plotone{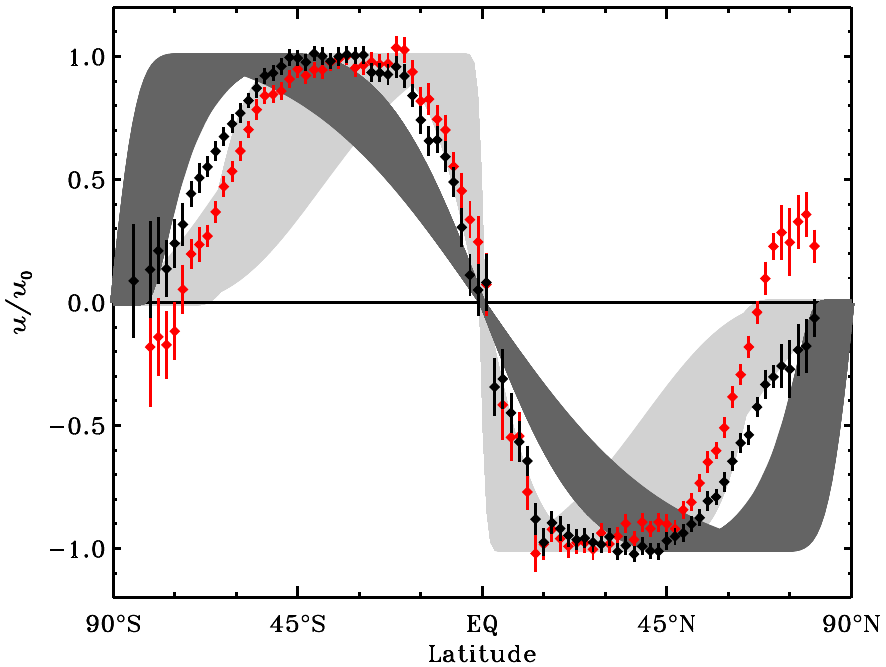}
   \caption{Observed and modeled latitudinal profiles of surface 
   meridional flow. 
   The dark gray band indicates the range of acceptable profiles in the 
   W21$\times$8-11 optimal solution of \S~\ref{s_optsol}, while the pale 
   gray band indicates the acceptable range obtained in Paper~I by 
   fitting the full synoptic magnetograms. 
   The solid dots and error bars are the Doppler measurements of
   \citet{Ulrich2010} for cycles 22 (red) and 23 (black).
   }
   \label{f_mercirc}
\end{figure}

We also carried out optimization runs in which the parameters defining 
the latitudinal dependence of the meridional flow (via 
Equations~(\ref{eq_mercirca})---(\ref{eq_mercircb})) are constrained 
to a narrower range of acceptable values, corresponding to the best-fit 
surface flux transport solution obtained in Paper~I by fitting actual 
synoptic magnetograms, rather than just the spatiotemporal distributions 
of \ac{BMR} emergences. 
The best-fit solution from such an optimization reaches only 
$C\simeq 0.86$, which is much less satisfactory than the $C=0.94$ 
best-fit solution. 
More worrisome is the fact that the surface meridional flow for the 
best-fit solution and error bars of Table~\ref{t_param}, plotted in 
Figure~\ref{f_mercirc} (dark gray band), provides a rather poor fit to 
the Doppler observations of \citet{Ulrich2010}, which lie mostly
outside the range of acceptable solutions from the optimization run. 
The best-fit profile of Paper~I did much better in this respect 
(reproduced herein as the pale gray band in Figure~\ref{f_mercirc}).

This suggests some incompatibility between the optimization of the 
\ac{SFT} model relative to surface magnetograms and the optimization of 
the coupled \ac{SFT}--\ac{FTD} model relative to the shape of the 
sunspot butterfly diagram. 
The W21$\times$8-11 optimal solution of \S~\ref{s_optsol} still lies 
within the surface-optimized ranges for the maximum meridional flow 
amplitude $u_0$, the surface diffusivity $\eta_R$, and the exponential 
decay time $\tau_R$ obtained in paper I, while the parameters $q$, $v$ 
and $w$ (see Equation~(\ref{eq_mercircb})), setting the latitudinal 
dependence of the stream function, do not. 
Interestingly, despite significant variations in latitudinal profiles, 
all acceptable solutions $(C\geq0.92$) have a peak equatorward 
meridional flow speed of $\unit{6-7}{\mps}$ near the base of the 
circulation cell; 
this is consistent with the deep meridional flow setting the cycle 
period in these dynamo solutions, which leads to a very tight constraint 
when fitting the butterfly diagram.

The analytic form adopted here for the meridional flow stream function 
is of course extremely simple: steady and separable in $r$ and $\theta$, 
which enforces the same latitudinal dependence at all depths, and
defining a single flow cell per meridional quadrant. 
What our butterfly diagram-based goodness-of-fit measure thus constrains 
is primarily the flow at the base of the convection zone. 
The misfit with the results from purely surface optimization suggests 
that the internal flow is more complex than the single-cell profile used 
here. 
Indeed, the recent helioseismic inversions of \citet{Zhao2013} and 
\citet{Schad2013} suggest multiple cells in radius, which is known to 
have a large impact on the operation of flux transport dynamos 
(e.g., \citealt{Jouve2007}). 
The dynamo modeling work of \citet{Hazra2014} indicates, however, that 
provided additional transport processes such as turbulent diffusion 
and/or pumping can couple the surface and base of the convection zone, 
solar-like butterfly diagrams can be produced as long as an equatorward 
flow is present at or immediately beneath the base of the convection 
zone (see also \citealt{Jiang2013}).

Another physical inconsistency of the W21$\times$8-11 optimal solution 
is the meridional flow's deep penetration below the base of the 
convection zone. 
This is known to be conducive to the production of solar-like butterfly 
diagrams (e.g., \citealt{Nandy2002}), but unlikely on dynamical grounds 
\citep{Gilman2004}, and delicate to reconcile with observed solar light 
element abundances \citep{Charbonneau2007a}. 
Finally, both observations \citep{Ulrich2010} and numerical simulations 
\citep{Passos2012} suggest that the meridional flow may undergo 
systematic temporal variations in the course of the cycle, 
presumably driven by the cycling magnetic field. 
Such effects are {\it a priori} excluded from the meridional flow 
parametrization used here.

All these incompatibilities and inconsistencies most likely reflect, 
at least in part, the specific choices made for the parametrization 
of the meridional flow profile. 
An interesting possibility would be to use our \ac{GA}-based 
fitting technique to invert a spatially-resolved discretization of 
the internal meridional flow from the sunspot butterfly diagram. 
Such a method, dubbed genetic forward modeling, has already been used 
successfully to infer the rotational profile of the deep solar core 
from low-$\ell$ rotational frequency splittings 
(see \citealt{Charbonneau1998}).

Genetic forward modeling could also be used to invert stability 
diagrams for the emergence of \acp{BMR}. 
Our best-fit emergence function has $a=0$ in 
Equation~(\ref{eq_fitfunc}), implying that the emergence probability is 
primarily set by the strength of the toroidal magnetic component, 
in agreement with the idea that sunspots form from axisymmetric toroidal 
magnetic flux ropes located at or near the base of the convection zone. 
However, our eruption threshold of $\simeq\unit{200}{\gauss}$ is rather 
low, even if some level of amplification is expected in forming a 
compact flux rope from a diffuse magnetic field. 
There is clearly room for improvement in this model component.

\section{Conclusions}


In this paper we have described a new solar cycle model based on the 
\acl{BL} mechanism of poloidal field regeneration through the surface 
decay of active regions. 
This new model is based on the coupling of a conventional 
latitude--longitude simulation of surface magnetic flux evolution (as 
described in Paper~I), coupled to an equally conventional axisymmetric 
kinematic flux transport dynamo model defined in a meridional plane 
(closely following \citealt{Charbonneau2005}). 
The novelty lies in the coupling between these to model components: the 
surface flux evolution simulation provides the source term of the 
internal dynamo through the surface boundary condition; while the 
internal dynamo provides the magnetic flux emergence, in the form of 
pseudo-sunspot bipolar pairs, that act as a source in the surface 
magnetic flux simulation. 
The properties of these synthetic bipolar pairs ---flux distribution, 
component separation, tilt angles, etc--- are tailored to reflect 
observed statistical properties of real sunspots and active regions, as 
documented in Paper~I (Appendix).

The other key aspect of the coupling is the emergence function, which 
controls the probability of bipole emergence as a function of the 
spatiotemporal distribution of the deep-seated magnetic field 
produced by the dynamo component of the coupled model. 
The emergence probability is assumed to scale linearly with this 
emergence function, with the proportionality constant acting as the 
dynamo number for the full coupled model.

The coupled model involves a number of parameters and functionals 
that cannot be set from first principle, and thus must be optimized 
to provide the best possible fit to solar observations. 
We opted to carry out this optimization task through a 
genetic algorithm-based
maximization of the fit between the spatiotemporal distribution of 
sunspot emergences (butterfly diagram) as produced by the model, and the
cycle 21 emergence data of \citet{Wang1989-0}. 
This scheme returns not only a globally optimal solution, but also 
Monte Carlo-like error estimates on best-fit parameters values.

The magnetic cycles generated by this dynamo model are intrinsically 
non-steady, due primarily to the large statistical scatter about the 
mean East--West tilt pattern of \acp{BMR} (as embodied in Joy's Law). 
This is expected, since the axial dipole component of the bipolar pair 
is determined by this tilt. 
As a consequence, a critical dynamo number can only be defined in a 
statistical sense.

A quenching parametrization of the mean tilt angle based on the 
strength of the internal magnetic field readily stabilizes the mean 
cycle amplitude, but large fluctuations about this mean nonetheless 
persist. 
Such a quenching is consistent with the modeling of the buoyant rise of 
thin magnetic flux tubes 
(see \citealt{Fan2009}, \S~5.1.2, and references therein), and, 
at the relatively mild level taking place in our dynamo model, does 
not conflict with extant observational analyses 
(see \citealt{Pevtsovetal2014}). 
One consequence of tilt quenching is that a very high amplitude cycle 
tend to be followed by a lower-than-average cycle. 
This alternation would tend to amplify over time were it not for the 
stabilizing effect of the linear sink term used in 
Equation~(\ref{eq_surftrans}) with $\tau_R=\unit{10}{\years}$.
Very low amplitude cycles can also be produced by unfavorable emergence 
patterns, which then lead to persistently low amplitudes in subsequent 
cycles, with slow recovery to normal amplitude values. 

Even though the amplitude of successive simulated cycles are strongly 
affected by the specific stochastic realization of flux, separation and 
tilts in the course of a given cycle, even in the linear regime the 
cycle period is largely insensitive to the value of the dynamo number. 
The magnetic cycle is also characterized by good hemispheric coupling, 
in terms of both hemispheric cycle amplitude and timing of hemispheric 
minima/maxima.

As a descriptive representation of the observed solar cycle, the model 
reproduces a number of well-known features. 
The dipole peaks at or slightly before the time of pseudo-sunspot cycle 
minimum, and its amplitude shows no correlation with the maximum 
pseudo-sunspot number of the ending cycle. 
This is a direct consequence of the strong stochasticity introduced by 
the realization of tilt patterns throughout the cycle, which is the 
primary source of cycle amplitude fluctuations. 
However, the model reproduces the observed positive correlation between 
dipole strength at cycle minimum and the amplitude of the subsequent 
pseudo-sunspot cycle. 
This indicates that, as in the real Sun, the dipole moment generated 
in the model is a good precursor of cycle amplitude.

Room for improvement certainly remains. 
The model fails to reproduce the observed moderate anticorrelation 
between cycle amplitude and duration, yielding instead a very weak 
positive correlation between these two quantities. 
While a few extant kinematic flux transport dynamo models do better in 
this respect (e.g., \citealt{Karak2011}), another possibility is that 
the origin of this pattern is to be found in dynamical effects, namely 
the magnetic backreaction on large-scale flows. 
The recent analyses of \citet{Passos2012} suggest that an increase in 
the speed of the deep equatorward meridional flow may indeed be driven 
by a higher-than-average large-scale magnetic field, which in 
advection-dominated flux transport dynamos would be expected to lead 
to a proportional reduction in cycle period 
(see, e.g., \citealt{Dikpati1999}).

The long timescale behavior of the simulated cycles also shows some
interesting features, some solar-like and others less so. 
The model produces a very stable cycle period of $9.5-11$~years, 
but no well-defined low-frequency spectral peaks that could be 
associated with Gleissberg-like long periodicities. 
The model does produce occasional Dalton-minimum-like periods of 
successive low amplitude cycles, and can also spontaneously shut down 
the cycle and enter a non-cycling grand-minima-like state, through an 
unfavorable stochastic pattern of bipolar pseudo-sunspot emergences in 
the course of a cycle. 
This is a relatively common occurrence for a simulation using the 
best-fit parameter values obtained in \S~\ref{s_calib}: more than one 
half of simulations initialized with distinct random seeds were found 
to undergo shutdown at some point during a 100-cycle long time span.

In subsequent papers in this series we will investigate cycle 
fluctuation patterns in greater detail, and quantify the occurrence 
statistics of Dalton-like minima. 
The few such events found so far in our extant simulation runs suggest 
that entry into these failed minima is rapid, from one cycle to 
the next, while recovery to average cycle amplitudes is more gradual. 
We also plan to add a weak turbulent alpha-effect in the convective
envelope portion of the domain, and investigate whether this can pull
the model out of a shutdown state, as existing simulations have already 
suggested (e.g., \citealt{Ossendrijver2000}; \citealt{Karak2013}; 
\citealt{Hazra2014b}).

Because it includes an explicit, spatially-resolved representation
of the solar ``surface'', the $2\times2$D solar cycle model presented 
here is ideally suited for providing synthetic data for coronal 
magnetic field reconstructions, as well as for assimilation of 
magnetographic data towards solar cycle forecasting. 
The results presented in this paper indicate that an accurate 
determination of the tilt angles of individual emerging bipolar sunspot 
pairs will be a critical element of such latter endeavor.



\begin{acronym}
   \acro{AR}{active region}
   \acro{BL}{Babcock--Leighton}
   \acro{BMR}{bipolar magnetic region}
   \acro{FTD}{flux transport dynamo}
   \acro{GA}{Genetic Algorithm}
   \acro{MHD}{magnetohydrodynamics}
   \acro{PDF}{probability distribution function}
   \acro{RMS}{root-mean-square}
   \acro{SFT}{surface flux transport}
   \acro{WS}{Wang and Sheeley}
\end{acronym}



\acknowledgments
We wish to thank
Yi-Ming Wang and Neil R. Sheeley, Jr. for kindy providing us with their 
comprehensive database of bipolar emergences for cycle 21
and Roger Ulrich for his compilation of latitudinal flow measurements 
and error estimates.
This research was funded by
a graduate research felloship of the Fonds de Recherche du Qu\'ebec 
Nature et Technologies (A. L.) and the Discovery Grant Program (P. C.)
of the Natural Science and Engineering Research Council of Canada.
Calculations were performed on Calcul Qu\'ebec's computing facilities,
a member of Compute Canada consortium.







\bibliographystyle{apj}
\bibliography{references}

\newcommand{\SortNoop}[1]{}
\begin{thebibliography}{}
\expandafter\ifx\csname natexlab\endcsname\relax\def\natexlab#1{#1}\fi

\bibitem[{{Babcock}(1959)}]{Babcock1959}
{Babcock}, H.~D. 1959, \apj, 130, 364

\bibitem[{{Babcock}(1961)}]{Babcock1961}
{Babcock}, H.~W. 1961, \apj, 133, 572

\bibitem[{{Babcock} \& {Babcock}(1955)}]{Babcock1955}
{Babcock}, H.~W., \& {Babcock}, H.~D. 1955, \apj, 121, 349

\bibitem[{{Baumann} {et~al.}(2006){Baumann}, {Schmitt}, \&
  {Sch{\"u}ssler}}]{Baumann2006}
{Baumann}, I., {Schmitt}, D., \& {Sch{\"u}ssler}, M. 2006, \aap, 446, 307

\bibitem[{{Baumann} {et~al.}(2004){Baumann}, {Schmitt}, {Sch{\"u}ssler}, \&
  {Solanki}}]{Baumann2004}
{Baumann}, I., {Schmitt}, D., {Sch{\"u}ssler}, M., \& {Solanki}, S.~K. 2004,
  \aap, 426, 1075

\bibitem[{{Bogdan} {et~al.}(1988){Bogdan}, {Gilman}, {Lerche}, \&
  {Howard}}]{Bogdan1988}
{Bogdan}, T.~J., {Gilman}, P.~A., {Lerche}, I., \& {Howard}, R. 1988, \apj,
  327, 451

\bibitem[{{Burnett}(1987)}]{Burnett1987}
{Burnett}, D.~S. 1987, Finite Element Analysis: From Concepts to Applications
  (Reading, Massachusetts: Addison-Wesley Pub. Co.)

\bibitem[{{Caligari} {et~al.}(1995){Caligari}, {Moreno-Insertis}, \&
  {Schussler}}]{Caligari1995}
{Caligari}, P., {Moreno-Insertis}, F., \& {Schussler}, M. 1995, \apj, 441, 886

\bibitem[{{Cameron} \& {Sch{\"u}ssler}(2015)}]{CameronSchussler2015}
{Cameron}, R., \& {Sch{\"u}ssler}, M. 2015, Science, 347, 1333

\bibitem[{Charbonneau(2002)}]{Charbonneau2002b}
Charbonneau, P. 2002, NCAR Tech. Note, NCAR/TN-451+STR (Boulder: National
  Center for Atmospheric Research), 1

\bibitem[{{Charbonneau}(2007)}]{Charbonneau2007a}
{Charbonneau}, P. 2007, Advances in Space Research, 39, 1661

\bibitem[{{Charbonneau}(2010)}]{Charbonneau2010}
---. 2010, Living Reviews in Solar Physics, 7, 3

\bibitem[{{Charbonneau}(2014)}]{Charbonneau2014}
---. 2014, \araa, 52, 251

\bibitem[{{Charbonneau} {et~al.}(1999){Charbonneau}, {Christensen-Dalsgaard},
  {Henning}, {Larsen}, {Schou}, {Thompson}, \& {Tomczyk}}]{Charbonneau1999}
{Charbonneau}, P., {Christensen-Dalsgaard}, J., {Henning}, R., {et~al.} 1999,
  \apj, 527, 445

\bibitem[{Charbonneau \& Knapp(1995)}]{Charbonneau1995}
Charbonneau, P., \& Knapp, B. 1995, NCAR Tech. Note, NCAR/TN-418+IA (Boulder:
  National Center for Atmospheric Research), 1

\bibitem[{{Charbonneau} {et~al.}(2005){Charbonneau}, {St-Jean}, \&
  {Zacharias}}]{Charbonneau2005}
{Charbonneau}, P., {St-Jean}, C., \& {Zacharias}, P. 2005, \apj, 619, 613

\bibitem[{{Charbonneau} {et~al.}(1998){Charbonneau}, {Tomczyk}, {Schou}, \&
  {Thompson}}]{Charbonneau1998}
{Charbonneau}, P., {Tomczyk}, S., {Schou}, J., \& {Thompson}, M.~J. 1998, \apj,
  496, 1015

\bibitem[{{Choudhuri} {et~al.}(2007){Choudhuri}, {Chatterjee}, \&
  {Jiang}}]{Choudhuri2007}
{Choudhuri}, A.~R., {Chatterjee}, P., \& {Jiang}, J. 2007, Physical Review
  Letters, 98, 131103

\bibitem[{{Choudhuri} {et~al.}(1995){Choudhuri}, {Sch{\"u}ssler}, \&
  {Dikpati}}]{Choudhuri1995}
{Choudhuri}, A.~R., {Sch{\"u}ssler}, M., \& {Dikpati}, M. 1995, \aap, 303, L29+

\bibitem[{{Dasi-Espuig} {et~al.}(2010){Dasi-Espuig}, {Solanki}, {Krivova},
  {Cameron}, \& {Pe{\~n}uela}}]{DasiEspuig2010}
{Dasi-Espuig}, M., {Solanki}, S.~K., {Krivova}, N.~A., {Cameron}, R., \&
  {Pe{\~n}uela}, T. 2010, \aap, 518, A7

\bibitem[{{Dikpati}(2011)}]{Dikpati2011}
{Dikpati}, M. 2011, \apj, 733, 90

\bibitem[{{Dikpati} \& {Charbonneau}(1999)}]{Dikpati1999}
{Dikpati}, M., \& {Charbonneau}, P. 1999, \apj, 518, 508

\bibitem[{{Dikpati} {et~al.}(2006){Dikpati}, {de Toma}, \&
  {Gilman}}]{Dikpati2006a}
{Dikpati}, M., {de Toma}, G., \& {Gilman}, P.~A. 2006, \grl, 33, 5102

\bibitem[{{Dikpati} \& {Gilman}(2001)}]{Dikpati2001}
{Dikpati}, M., \& {Gilman}, P.~A. 2001, \apj, 559, 428

\bibitem[{{Dikpati} \& {Gilman}(2007)}]{Dikpati2007}
---. 2007, \solphys, 241, 1

\bibitem[{{D'Silva} \& {Choudhuri}(1993)}]{DSilva1993}
{D'Silva}, S., \& {Choudhuri}, A.~R. 1993, \aap, 272, 621

\bibitem[{{Durney}(1995)}]{Durney1995}
{Durney}, B.~R. 1995, \solphys, 160, 213

\bibitem[{{Fan}(2009)}]{Fan2009}
{Fan}, Y. 2009, Living Reviews in Solar Physics, 6, 4

\bibitem[{{Fan} \& {Fang}(2014)}]{FanFang2014}
{Fan}, Y., \& {Fang}, F. 2014, \apj, 789, 35

\bibitem[{{Ferriz-Mas} {et~al.}(1994){Ferriz-Mas}, {Schmitt}, \&
  {Schuessler}}]{FerrizMas1994}
{Ferriz-Mas}, A., {Schmitt}, D., \& {Schuessler}, M. 1994, \aap, 289, 949

\bibitem[{{Gilman} \& {Miesch}(2004)}]{Gilman2004}
{Gilman}, P.~A., \& {Miesch}, M.~S. 2004, \apj, 611, 568

\bibitem[{{Hale} {et~al.}(1919){Hale}, {Ellerman}, {Nicholson}, \&
  {Joy}}]{Hale1919}
{Hale}, G.~E., {Ellerman}, F., {Nicholson}, S.~B., \& {Joy}, A.~H. 1919, \apj,
  49, 153

\bibitem[{{Hazra} {et~al.}(2014{\natexlab{a}}){Hazra}, {Karak}, \&
  {Choudhuri}}]{Hazra2014}
{Hazra}, G., {Karak}, B.~B., \& {Choudhuri}, A.~R. 2014{\natexlab{a}}, \apj,
  782, 93

\bibitem[{{Hazra} {et~al.}(2014{\natexlab{b}}){Hazra}, {Passos}, \&
  {Nandy}}]{Hazra2014b}
{Hazra}, S., {Passos}, D., \& {Nandy}, D. 2014{\natexlab{b}}, \apj, 789, 5

\bibitem[{{Howard}(1991)}]{Howard1991}
{Howard}, R.~F. 1991, \solphys, 136, 251

\bibitem[{{Jiang} {et~al.}(2013){Jiang}, {Cameron}, {Schmitt}, \& {I{\c
  s}{\i}k}}]{Jiang2013}
{Jiang}, J., {Cameron}, R.~H., {Schmitt}, D., \& {I{\c s}{\i}k}, E. 2013, \aap,
  553, A128

\bibitem[{{Jouve} \& {Brun}(2007)}]{Jouve2007}
{Jouve}, L., \& {Brun}, A.~S. 2007, \aap, 474, 239

\bibitem[{{Karak} \& {Choudhuri}(2011)}]{Karak2011}
{Karak}, B.~B., \& {Choudhuri}, A.~R. 2011, \mnras, 410, 1503

\bibitem[{{Karak} \& {Choudhuri}(2013)}]{Karak2013}
---. 2013, Research in Astronomy and Astrophysics, 13, 1339

\bibitem[{{Karak} {et~al.}(2014){Karak}, {Jiang}, {Miesch}, {Charbonneau}, \&
  {Choudhuri}}]{Karak2014}
{Karak}, B.~B., {Jiang}, J., {Miesch}, M.~S., {Charbonneau}, P., \&
  {Choudhuri}, A.~R. 2014, \ssr, 186, 561

\bibitem[{{Leighton}(1964)}]{Leighton1964}
{Leighton}, R.~B. 1964, \apj, 140, 1547

\bibitem[{{Lemerle} {et~al.}(2015){Lemerle}, {Charbonneau}, \&
  {Carignan-Dugas}}]{Lemerle2015}
{Lemerle}, A., {Charbonneau}, P., \& {Carignan-Dugas}, A. 2015, \apj, 810, 78

\bibitem[{{McClintock} \& {Norton}(2013)}]{McClintockNorton2013}
{McClintock}, B.~H., \& {Norton}, A.~A. 2013, \solphys, 287, 215

\bibitem[{{Miesch} \& {Dikpati}(2014)}]{Miesch2014}
{Miesch}, M.~S., \& {Dikpati}, M. 2014, \apjl, 785, L8

\bibitem[{{Miesch} \& {Teweldebirhan}(2015)}]{Miesch2016}
{Miesch}, M.~S., \& {Teweldebirhan}, K. 2015, ArXiv e-prints, arXiv:1511.03613

\bibitem[{{Mu{\~n}oz-Jaramillo} {et~al.}(2013){Mu{\~n}oz-Jaramillo},
  {Dasi-Espuig}, {Balmaceda}, \& {DeLuca}}]{Munoz2013}
{Mu{\~n}oz-Jaramillo}, A., {Dasi-Espuig}, M., {Balmaceda}, L.~A., \& {DeLuca},
  E.~E. 2013, \apjl, 767, L25

\bibitem[{{Mu{\~n}oz-Jaramillo} {et~al.}(2010){Mu{\~n}oz-Jaramillo}, {Nandy},
  {Martens}, \& {Yeates}}]{Munoz2010}
{Mu{\~n}oz-Jaramillo}, A., {Nandy}, D., {Martens}, P.~C.~H., \& {Yeates}, A.~R.
  2010, \apjl, 720, L20

\bibitem[{{Nandy} \& {Choudhuri}(2001)}]{Nandy2001}
{Nandy}, D., \& {Choudhuri}, A.~R. 2001, \apj, 551, 576

\bibitem[{{Nandy} \& {Choudhuri}(2002)}]{Nandy2002}
---. 2002, Science, 296, 1671

\bibitem[{{Nelson} {et~al.}(2013){Nelson}, {Brown}, {Brun}, {Miesch}, \&
  {Toomre}}]{Nelson2013}
{Nelson}, N.~J., {Brown}, B.~P., {Brun}, A.~S., {Miesch}, M.~S., \& {Toomre},
  J. 2013, \apj, 762, 73

\bibitem[{{Nelson} {et~al.}(2014){Nelson}, {Brown}, {Sacha Brun}, {Miesch}, \&
  {Toomre}}]{Nelson2014}
{Nelson}, N.~J., {Brown}, B.~P., {Sacha Brun}, A., {Miesch}, M.~S., \&
  {Toomre}, J. 2014, \solphys, 289, 441

\bibitem[{{Ossendrijver}(2000)}]{Ossendrijver2000}
{Ossendrijver}, M.~A.~J.~H. 2000, \aap, 359, 1205

\bibitem[{{Parker}(1955)}]{Parker1955}
{Parker}, E.~N. 1955, \apj, 122, 293

\bibitem[{{Passos} \& {Charbonneau}(2014)}]{Passos2014b}
{Passos}, D., \& {Charbonneau}, P. 2014, \aap, 568, A113

\bibitem[{{Passos} {et~al.}(2012){Passos}, {Charbonneau}, \&
  {Beaudoin}}]{Passos2012}
{Passos}, D., {Charbonneau}, P., \& {Beaudoin}, P. 2012, \solphys, 279, 1

\bibitem[{{Passos} {et~al.}(2014){Passos}, {Nandy}, {Hazra}, \&
  {Lopes}}]{Passosetal2014}
{Passos}, D., {Nandy}, D., {Hazra}, S., \& {Lopes}, I. 2014, \aap, 563, A18

\bibitem[{{Pevtsov} {et~al.}(2014){Pevtsov}, {Berger}, {Nindos}, {Norton}, \&
  {van Driel-Gesztelyi}}]{Pevtsovetal2014}
{Pevtsov}, A.~A., {Berger}, M.~A., {Nindos}, A., {Norton}, A.~A., \& {van
  Driel-Gesztelyi}, L. 2014, \ssr, 186, 285

\bibitem[{{Schad} {et~al.}(2013){Schad}, {Timmer}, \& {Roth}}]{Schad2013}
{Schad}, A., {Timmer}, J., \& {Roth}, M. 2013, \apjl, 778, L38

\bibitem[{{Schmitt}(1987)}]{Schmitt1987}
{Schmitt}, D. 1987, \aap, 174, 281

\bibitem[{{Schrijver} {et~al.}(2002){Schrijver}, {De Rosa}, \&
  {Title}}]{Schrijver2002}
{Schrijver}, C.~J., {De Rosa}, M.~L., \& {Title}, A.~M. 2002, \apj, 577, 1006

\bibitem[{{Sch{\"u}ssler} {et~al.}(1994){Sch{\"u}ssler}, {Caligari},
  {Ferriz-Mas}, \& {Moreno-Insertis}}]{Schussler1994}
{Sch{\"u}ssler}, M., {Caligari}, P., {Ferriz-Mas}, A., \& {Moreno-Insertis}, F.
  1994, \aap, 281, L69

\bibitem[{{Snodgrass}(1983)}]{Snodgrass1983}
{Snodgrass}, H.~B. 1983, \apj, 270, 288

\bibitem[{{Stenflo} \& {Kosovichev}(2012)}]{Stenflo2012}
{Stenflo}, J.~O., \& {Kosovichev}, A.~G. 2012, \apj, 745, 129

\bibitem[{{Ulrich}(2010)}]{Ulrich2010}
{Ulrich}, R.~K. 2010, \apj, 725, 658

\bibitem[{{van Ballegooijen} \& {Choudhuri}(1988)}]{vanBalle1988}
{van Ballegooijen}, A.~A., \& {Choudhuri}, A.~R. 1988, \apj, 333, 965

\bibitem[{{Wang} {et~al.}(2002{\natexlab{a}}){Wang}, {Lean}, \&
  {Sheeley}}]{Wang2002a}
{Wang}, Y.-M., {Lean}, J., \& {Sheeley}, Jr., N.~R. 2002{\natexlab{a}}, \apjl,
  577, L53

\bibitem[{{Wang} {et~al.}(1989){Wang}, {Nash}, \& {Sheeley}}]{Wang1989}
{Wang}, Y.-M., {Nash}, A.~G., \& {Sheeley}, Jr., N.~R. 1989, Science, 245, 712

\bibitem[{{Wang} \& {Sheeley}(1989)}]{Wang1989-0}
{Wang}, Y.-M., \& {Sheeley}, Jr., N.~R. 1989, \solphys, 124, 81

\bibitem[{{Wang} \& {Sheeley}(1991)}]{Wang1991}
---. 1991, \apj, 375, 761

\bibitem[{{Wang} {et~al.}(2002{\natexlab{b}}){Wang}, {Sheeley}, \&
  {Lean}}]{Wang2002b}
{Wang}, Y.-M., {Sheeley}, Jr., N.~R., \& {Lean}, J. 2002{\natexlab{b}}, \apj,
  580, 1188

\bibitem[{{Weber} {et~al.}(2011){Weber}, {Fan}, \& {Miesch}}]{Weber2011}
{Weber}, M.~A., {Fan}, Y., \& {Miesch}, M.~S. 2011, \apj, 741, 11

\bibitem[{{Yeates} \& {Mu{\~n}oz-Jaramillo}(2013)}]{YeatesMunoz2013}
{Yeates}, A.~R., \& {Mu{\~n}oz-Jaramillo}, A. 2013, \mnras, 436, 3366

\bibitem[{{Zhao} {et~al.}(2013){Zhao}, {Bogart}, {Kosovichev}, {Duvall}, \&
  {Hartlep}}]{Zhao2013}
{Zhao}, J., {Bogart}, R.~S., {Kosovichev}, A.~G., {Duvall}, Jr., T.~L., \&
  {Hartlep}, T. 2013, \apjl, 774, L29

\end{thebibliography}

\clearpage

\end{document}